\documentclass[twocolumn,times]{aastex62}

\usepackage{enumitem}
\usepackage{float}
\usepackage{makecell}
\usepackage{rotating}
\graphicspath{{./}{figures/}}

\shorttitle{Sensitivity of ${^{44}}$Ti and ${^{56}}$Ni production  in CCSN shock-driven nucleosynthesis to reaction rates}
\shortauthors{Subedi, Meisel, \& Merz}

\begin{document}

\title{Sensitivity of ${^{44}}$Ti and ${^{56}}$Ni production in core collapse supernova shock-driven nucleosynthesis to nuclear reaction rate variations}

\author{Shiv K. Subedi}
\altaffiliation{Affiliated with the Joint Institute for Nuclear Astrophysics--Center \\ for the Evolution of the Elements}
\affiliation{Institute of Nuclear \& Particle Physics, Department of Physics \& Astronomy, Ohio University, Athens, Ohio 45701, USA}
\email{ss383615@ohio.edu, meisel@ohio.edu}

\author[0000-0002-8403-8879]{Zach Meisel}
\altaffiliation{Affiliated with the Joint Institute for Nuclear Astrophysics--Center \\ for the Evolution of the Elements}
\affiliation{Institute of Nuclear \& Particle Physics, Department of Physics \& Astronomy, Ohio University, Athens, Ohio 45701, USA}

\author{Grant Merz}
\altaffiliation{Affiliated with the Joint Institute for Nuclear Astrophysics--Center \\ for the Evolution of the Elements}
\affiliation{Institute of Nuclear \& Particle Physics, Department of Physics \& Astronomy, Ohio University, Athens, Ohio 45701, USA}



\begin{abstract}
Recent observational advances have enabled high resolution mapping of ${^{44}}$Ti in core-collapse supernova (CCSN) remnants. Comparisons between observations and models provide stringent constraints on the CCSN mechanism. However, past work has identified several uncertain nuclear reaction rates that influence ${^{44}}$Ti and ${^{56}}$Ni production in post-processing model calculations.  We evolved one dimensional models of $15~M_{\odot}$, $18~M_{\odot}$, $22~M_{\odot}$ and $25~M_{\odot}$ stars from zero-age main sequence through CCSN using {\tt MESA} (Modules for Experiments in Stellar Astrophysics) and investigated the previously identified reaction rate sensitivities of ${^{44}}$Ti and ${^{56}}$Ni production. We tested the robustness of our results by making various assumptions about the CCSN explosion energy and mass-cut. We found a number of reactions that have a significant impact on the nucleosynthesis of ${^{44}}$Ti and ${^{56}}$Ni, particularly for lower progenitor masses. Notably, the reaction rates
$^{13}{\rm N}(\alpha,p)^{16}{\rm O}$, $^{17}{\rm F}(\alpha,p)^{20}{\rm Ne}$,  $^{52}{\rm Fe}(\alpha,p)^{55}{\rm Co}$, $^{56}{\rm Ni}(\alpha,p)^{59}{\rm Cu}$, $^{57}{\rm Ni}(n,p)^{57}{\rm Co}$, $^{56}{\rm Co}(p,n)^{56}{\rm Ni}$,
$^{39}{\rm K}(p,\gamma)^{40}{\rm Ca}$,
$^{47}{\rm V}(p,\gamma)^{48}{\rm Cr}$,  
$^{52}{\rm Mn}(p,\gamma)^{53}{\rm Fe}$,
$^{57}{\rm Co}(p,\gamma)^{58}{\rm Ni}$,  and $^{39}{\rm K}(p,\alpha)^{36}{\rm Ar}$
are influential for a large number of model conditions. Furthermore, we found the list of influential reactions identified by previous post-processing studies of CCSN shock-driven nucleosynthesis is likely incomplete, motivating future larger-scale sensitivity studies.\\


\end{abstract}

\keywords{Astrophysical explosive burning, Explosive nucleosynthesis, Core-collapse supernovae}


\section{Introduction} \label{sec:intro}

Stars with initial mass $\gtrsim8~M_{\odot}$ undergo a core-collapse supernova (CCSN) explosion after core fuel exhaustion \citep{2005NatPh...1..147W}. The CCSN enriches the interstellar medium by releasing the isotopes synthesized throughout its life cycle and during CCSN nucleosynthesis. Among the ejected isotopes, ${^{44}}{\rm Ti}$ and ${^{56}}{\rm Ni}$ are produced near the boundary of the proto-neutron star remnant and the ejecta, and are thus studied to gain insight into details of the CCSN mechanism \citep{0004-637X-640-2-891,2007ApJ...664.1033Y,2012ApJ...749...91F,Fryer_2018}.

Observations of $^{44}{\rm Ti}$ in CCSN remnants are possible due to characteristic $\gamma$-rays at 67.9, 78.3, and 1157.0~keV emitted in the decay sequence ${^{44}}{\rm Ti}$ ($t_{1/2} = 59.1$~y) $\rightarrow{^{44}}{\rm Sc}$ ($t_{1/2} = 3.97$~h) $\rightarrow{^{44}}{\rm Ca}$ (stable)~\citep{2011NDS...112.2357C}.
$^{56}{\rm Ni}$, on the other hand, follows the much briefer decay sequence ${^{56}}{\rm Ni}$ ($t_{1/2} = 6.08$~d) $\rightarrow{^{56}}{\rm Co}$ ($t_{1/2} = 77.2$~d) $\rightarrow{^{56}}{\rm Fe}$~\citep{2011NDS...112.1513J}. As such, CCSN $^{56}{\rm Ni}$ production is generally inferred from observations of the remnant iron, e.g. for the Cassiopeia A (CasA) supernova of 1671~AD~\citep{Eriksen_2009}. 
%

Comparing such observational constraints to yields from CCSN model calculations offers the opportunity to constrain properties of the CCSN explosion, such as the explosion energy and duration, as well as the remnant mass ~\citep{1991ApJ...370..630A,0004-637X-640-2-891,2007ApJ...664.1033Y,Mull17,Fryer_2018,Sawada2019}. However, model calculation results for $^{44}{\rm Ti}$ and $^{56}{\rm Ni}$ yields have been shown to have significant sensitivities to variations in nuclear reaction rates~\citep{The_1998,Hoff99,Magkotsios_2010}. This is particularly problematic as many of the relevant reaction rates have poor experimental constraints. Nuclear physics measurements can be prioritized with the aid of sensitivity studies, whereby nuclear reaction rates are varied within an uncertainty factor and the impact on the model calculation results is assessed. 

Large scale investigations of nuclear reaction rate uncertainties impacting shock-driven nucleosynthesis in CCSN were previously performed by  \cite{The_1998} and \cite{Magkotsios_2010}. These pioneering works used analytic temperature-density trajectories in nucleosynthesis post-processing calculations, varying nuclear reaction rates by factors of /100 and $\times$100. They identified several nuclear reaction rates that can impact ${^{44}}{\rm Ti}$ and ${^{56}}{\rm Ni}$ nucleosynthesis for temperature and density trajectories approximating material heated by the expanding shock following a supernova core bounce. We build on prior work in the following ways: (1) We perform stellar evolution and CCSN calculations with one-dimensional models in an effort to focus on astrophysical conditions that are most relevant to comparisons with astronomical observations. (2) We vary nuclear reaction rates by realistic rate variation factors based on existing nuclear physics constraints in order to avoid focusing on cases which are already sufficiently constrained. (3) We cross-check whether the prioritized reaction rate lists resulting from prior studies are comprehensive. We note that similar work has been done by \citet{Hoff10} and \citet{Tur_2010}, but for a much smaller set of reaction rate variations.

Our paper is organized as follows. In Section \ref{sec:zams_ccsn} we discuss the model details for massive star evolution and subsequent CCSN. In Section \ref{sec:them_pro} we detail temperature and density evolution during CCSN for comparison with prior post-processing work. In Section \ref{sec:nuc_yield} we present the explosion details, nucleosynthetic yields of ${^{44}}{\rm Ti}$ and ${^{56}}{\rm Ni}$ from our baseline CCSN calculations, and compare to yields determined by observations and previous modeling efforts. In Section \ref{sec:rr_var} we present our list of varied reaction rates and adopted reaction rate variation factors. In Section \ref{sec:sen_study} we present nuclear physics sensitivities, explaining the rate impacts in terms of the nuclear reaction network. In Section \ref{sec:comp} we compare our results with the post-processing calculations of \citet{The_1998} and \citet{Magkotsios_2010}. We conclude in Section \ref{sec:conc}, with brief suggestions for future work involving astrophysics model calculations and nuclear physics experiments.

\section{{\tt MESA} Calculations} \label{sec:zams_ccsn}
We performed one-dimensional, multi-zone model calculations using {\tt MESA}\footnote{http://mesa.sourceforge.net}(Modules for Experiments in Stellar Astrophysics), which is an open-source code for modeling stellar evolution and stellar explosions \citep{2011ApJS..192....3P,2015ApJS..220...15P,2018ApJS..234...34P,2019arXiv190301426P}.
In order to perform CCSN, we ran our simulations in two different stages using {\tt MESA} version 7624, consisting of evolution to core collapse followed by the supernova explosion.

In stage one, we evolved non-rotating, solar metallicity ($Z= 0.02$) $M_{\rm prog}=15~M_{\odot}$, $18~M_{\odot}$, $22~M_{\odot}$ and $25~M_{\odot}$ progenitors from zero-age main sequence (ZAMS) 
to the onset of core-collapse, as this is within the suspected ZAMS mass range for progenitors of CCSN remnants with observed ${^{44}}$Ti~\citep{2002RMxAC..12...94P,0004-637X-640-2-891,0004-637X-842-1-13}.
Our stage one calculations are based on \citet{2016ApJS..227...22F}, using the 204 isotope network {\tt mesa\_204.net}\footnote{
\citet{Magkotsios_2010} found this network size is sufficient for the temperature and density phase-space and initial $Y_{e}$ relevant for our work.}, the JINA ReacLib reaction rate library version V2.0 2013-04-02~\citep{Cyburt_2010}, and the {\tt Dutch} wind scheme that is based on Refs.~\citep{1990A&A...231..134N,2000A&A...360..227N,2001A&A...369..574V,2009A&A...497..255G} with an efficiency scale factor $\eta= 0.8$~\citep{2001A&A...373..555M}. The spatial resolution resulted in the number of zones varying between $\sim450-1200$ zones, with the number of zones tending to increase in later stages of stellar evolution. The onset of core-collapse was defined as the time when any mass zone at the interior of the star exceeded an infall velocity of 1000 km\,s$^{-1}$~\citep{2016ApJS..227...22F}. 

Stage two took the final stellar model of stage one and carried this into the CCSN phase, following the procedure described by~\citet{2015ApJS..220...15P}, which is divided into four distinct steps: 
\begin{enumerate}
\item Run {\tt MESA} using {\tt inlist\_adjust}, where we adjust several control parameters for initiating CCSN simulations. The astrophysical parameters were chosen following~\citet{2016ApJS..227...22F}. Notable among them are absence of any prescription to mimic rotation and the choice of the {\tt Dutch} wind scheme with $\eta=0.8$.

\item Run {\tt MESA} using {\tt inlist\_remove\_core}, where we define and remove the stellar core. The outer mass coordinate of the core, i.e. the deepest zone of the model, is defined by the inner mass boundary $I_{\rm b}$, located at the mass coordinate just outside the Fe core (as specified in Table~\ref{tab:abund}).

\item Run {\tt MESA} using {\tt inlist\_edep}, which employs the ``thermal bomb'' mechanism~\citep[e.g.][]{2007ApJ...664.1033Y,2015ApJS..220...15P} to inject energy $E_{\rm inj}\sim\,10^{51}$ erg~\citep{2012ApJ...749...91F} into a thin mass shell $\Delta M_{\rm shell}=0.05M_{\odot}$ above $I_{\rm b}$ (the innermost $\sim$ 20 zones) over a time period $t_{\rm inj}= 20$~ms. As specified in Table~\ref{tab:abund}, we inject $E_{\rm inj} = 0.5$~foe\footnote{$1~{\rm foe}= 10^{51}$~erg} in $15~M_{\odot}$ and $18~M_{\odot}$ models; 1.32~foe in $15~M_{\odot}$, $18~M_{\odot}$, and $22~M_{\odot}$; and 3.5~foe in $22~M_{\odot}$ and $25~M_{\odot}$ ZAMS stars. These explosion energies were roughly based on the findings of \citet{2019MNRAS.482..351V,2012ApJ...749...91F,2015ApJS..220...15P,2014IAUS..296...27N}, where the specific energies were somewhat arbitrarily chosen in order to span a plausible range. 

We limited our ourselves to a single $t_{\rm inj}$ to keep the overall number of calculations manageable. Our choice of $t_{\rm inj}$ was motivated in part by the finding of~\citet{2012ApJ...749...91F} that most supernovae have explosion energies $E_{\rm exp}>10^{51}$ erg, which can be reached if the explosion occurs less than 250 ms after core bounce. Additionally, the analysis of \citet{Sawada2019} suggests short explosion timescales are more consistent with observations. The specific choice of $t_{\rm inj}=20$~ms corresponds to the fast-explosion time scale for thermal bomb models explored in \citet{2007ApJ...664.1033Y}. To explore the impact of this choice, we performed CCSN calculations for the $18~M_{\odot}$ model using $E_{\rm inj}$= 1.32 foe and $t_{\rm inj}=3, 20,$ and 200~ms, comparing the yield of $^{44}{\rm Ti}$ (see Section~\ref{sec:nuc_yield}). The 20~ms calculations were in closer agreement with the $^{44}{\rm Ti}$ yields of 3D models and observations, further motivating our choice. Similarly, we explored using $\Delta M_{\rm shell}=0.02\,M_{\odot}$, finding little impact on the $^{44}{\rm Ti}$ and $^{56}{\rm Ni}$ yields.

\item Run {\tt MESA} using {\tt inlist\_explosion}, which follows the thermodynamic and nucleosynthetic evolution of the stellar envelope due to the shockwave propagating outward from the thermal bomb energy deposition region. Nuclear reaction rates were varied in this step of the calculations.
\end{enumerate}

\section{Evolution of temperature and density during CCSN} \label{sec:them_pro}
As the shock wave powered by the thermal bomb propagates towards the stellar surface, it changes the thermodynamic conditions in the regions of the star that it sweeps through. We quantify the impact on thermodynamic conditions by comparing to analytic temperature-density trajectories, namely exponential and power-law, often used in post-processing studies~\citep[e.g.][]{The_1998,Magkotsios_2010}.
Both trajectories are characterized by a peak temperature $T_{0}$ and peak density $\rho_{0}$, followed by expansion and cooling with constant $T^{3}\rho^{-1}$~\citep{Magkotsios_2010}.

The exponential trajectory is based on \citet{1964ApJS....9..201F}, where material heated to $T_{0}$ and compressed to $\rho_{0}$ undergoes adiabatic expansion, and the expansion timescale $\tau$ is equal to the free-fall timescale:
\begin{equation}
\label{eqn:tau}
\tau = (24\pi G \rho_{0})^{-1/2} \approx 446(\rho_{0})^{-1/2}~{\rm s},
\end{equation}
where $G$ is Newton's gravitation constant. The associated temporal evolution of temperature and density are described by~\citet{Magkotsios_2010,2017ApJ...843....2H,Fryer_2018}
\begin{equation}
\label{eqn:adtemp}
    T(t) = T_{0}\exp{\Big(-\frac{t}{3\tau}\Big)} 
\qquad
\rho(t) = \rho_{0}\exp{\Big(-\frac{t}{\tau}\Big)}.
\end{equation}

The power-law profile is based on a constant-velocity (homologous) expansion, see e.g.~\citet{Magkotsios_2010,Fryer_2018}, where the temporal evolution of shock-heated material is described as
\begin{equation} 
\label{eqn:pow}
T(t) = \frac{T_{0}}{2t+1} \qquad \rho(t)= \frac{\rho_{0}}{(2t+1)^3} .
\end{equation}

To inform comparisons with prior post-processing studies using these analytic trajectories, we separately fit the temperature and density evolution of each zone of our models using Equations \ref{eqn:adtemp} and \ref{eqn:pow}. We found that the power-law profile provides a superior reproduction of our data. This is consistent with \citet{Fryer_2018}, who found the power-law describes their data particularly well once the shock-heated material drops out of nuclear statistical equilibrium (NSE).

 \begin{figure}[H]
 \epsscale{1.2}
\plotone{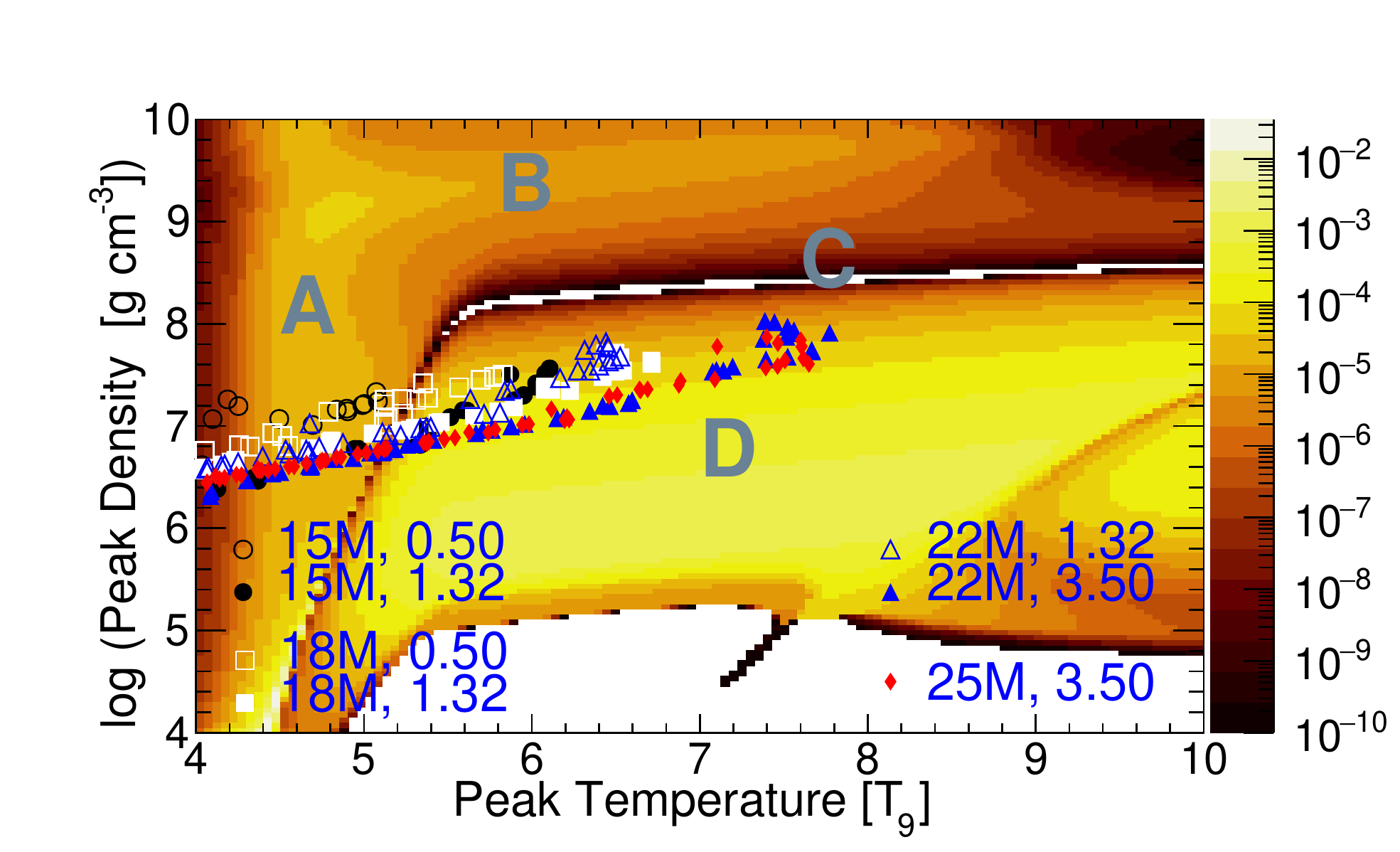}
\caption{Peak temperature and peak density phase space traversed by radial mass zones in {\tt MESA} for CCSN resulting from $M_{\rm prog}=15~M_{\odot}$, $18~M_{\odot}$, $22~M_{\odot}$, and $25~M_{\odot}$ models. As identified in \cite{Magkotsios_2010}, regions labelled in the plot are defined as A: Si-rich, B: Normal, C: QSE leakage, and D: $\alpha$-rich. The legend refers to $M_{\rm prog}$, followed by $E_{\rm inj}$ in $10^{51}$ erg. The color map indicates the $^{44}{\rm Ti}$ mass fraction $X({^{44}}{\rm Ti})$, as calculated for power law trajectories with $Y_{e} = 0.498$ by \cite{Magkotsios_2010}.}
\label{fig:temp_dens}
\end{figure}

\cite{Magkotsios_2010} studied the trends for ${^{44}}$Ti and ${^{56}}$Ni production in the $T_{0}$--$\rho_{0}$ plane for both exponential and power law profiles. Figure~\ref{fig:temp_dens} shows a color map of their $^{44}{\rm Ti}$ production across the $T_{0}$--$\rho_{0}$ phase-space for the power-law profile (Equation~\ref{eqn:pow}) with electron-fraction $Y_{e}=0.498$, which is close to the average $Y_{e}=0.498 - 0.499$ range that we observe across each of our progenitor stars.
For each $M_{\rm prog}$ and $E_{\rm inj}$ used in our CCSN calculations, we determined $T_{0}$ and $\rho_{0}$ for each mass zone, including these as data points in Figure~\ref{fig:temp_dens}. The figure includes labels for the four regions identified by \citet{Magkotsios_2010} that are characterized by specific nuclear burning patterns and govern the yield of ${^{44}}{\rm Ti}$. Region A corresponds to incomplete silicon burning. Region B corresponds to normal freeze-out from NSE, where the abundance is largely determined from the Q-values of the reactions~\citep{1973ApJS...26..231W,1994ARA&A..32..153M,The_1998,Hix_1999}. Region C corresponds to the chasm region, where nucleosynthesis is characterized by the flow of material from a low-$A$ quasistatic equilibium (QSE) cluster containing $^{44}{\rm Ti}$ to a high-$A$ cluster containing $^{56}{\rm Ni}$. Region D corresponds to the $\alpha$-rich freeze out~\citep{1973ApJS...26..231W}.

\citet{Magkotsios_2010} demonstrated that the yield of ${^{44}}{\rm Ti}$ in shock-driven nucleosynthesis depends sensitively on the location in the $T_{0}$--$\rho_{0}$ plane and the expansion profile followed by the shock-heated material. The reaction rate sensitivities reported by that work were therefore specified by phase-space region and expansion profile, including the impact on the topology of the phase space (e.g. chasm widening). Figure \ref{fig:temp_dens} shows that our $M_{\rm prog}=15~M_{\odot}$ $E_{\rm inj} = 0.5$~foe model is limited to lower values of $T_{0}$ and does not cross the chasm region (Region C), while any other combination of $E_{\rm inj}$ with progenitor mass $M_{\rm prog}$ extends to higher $T_{0}$ and $\rho_{0}$, into the $\alpha$-rich freeze-out (Region D). The increase in maximum $T_{0}$ and $\rho_{0}$ with increasing $M_{\rm prog}$ and $E_{\rm inj}$ is also apparent. 

\begin{table*}
\begin{center}
\caption{Key model properties and the resultant  $M({^{44}}{\rm Ti})$ and $M({^{56}}{\rm Ni})$ with $\Delta M_{\rm shell}= 0.05\,M_{\odot}$ for the baseline calculations performed in this work. ${\rm Fe}_{\rm core}$ is the mass of the iron core. $T_{0}$, Log$_{10}$($\rho_{0})$ and $Y_{e}$ represent corresponding values from $M_{4}$ to the mass coordinate with no oxygen-burning. All other properties are described in the text. \label{tab:abund} }
\begin{tabular}{c|c|c|c|c|c|c|c|c|c|c|c}


\hline \hline
$M_{\rm prog}$ & $E_{\rm inj}$  & $E_{\rm exp}$  & ${\rm Fe}_{\rm core}$ & $I_{\rm b}$ & $M_{\rm cut}$ & $T_{0}$ & Log$_{10}$($\rho_{0}$) & $Y_{e}$ & $M({^{44}} {\rm Ti})$ &  $M({^{56}} {\rm Ni})$ & $\frac{ M(^{44}\rm Ti)}{M(^{56}\rm Ni)}$\\ \newline
 [$M_{\odot}$] & [foe] & [foe] & [$M_{\odot}$] & [$M_{\odot}$] & [$M_{\odot}$] & [GK] & $\rho_{0}$[gm cm$^{-3}$] &  & $\times10^{-5}$ [$M_{\odot}$] & $\times10^{-2}$ [$M_{\odot}$] &  $\times10^{-5}$\\
\hline \hline
    & 0.50 & 0.24 & 1.40 & 1.45 & 1.450 &  &  &  & 1.22  &  9.0 &  13.9\\
    & 0.50 & 0.24 & 1.40 & 1.45 & 1.500 &  &   &  &  0.43 &   5.0 & 8.14 \\
15  & 0.50 & 0.24 & 1.40 & 1.45 & 1.600 & 4.848 - 2.483 & 7.120 - 5.962 & 0.497 - 0.499 & 0.08 &   0.3 & 23.9 \\
\cline{2-12}
   & 1.32 & 1.13 & 1.40 & 1.45 & 1.450 &  &  &  & 1.19 &   16.0 & 7.17  \\
   & 1.32 & 1.13 & 1.40 & 1.45 & 1.500 &  &  &  &  0.24 &   14.0 & 1.78 \\
   & 1.32 & 1.13 & 1.40 & 1.45 & 1.600 & 5.522 - 2.763 & 7.106 - 5.967  & 0.497 - 0.499 & 0.21 &    6.0 & 3.71 \\
\hline
    & 0.50 & 0.07 & 1.46 & 1.45 & 1.450 &  &  &  &  1.10 &   14.0 & 8.02 \\
    & 0.50 & 0.07 & 1.46 & 1.45 & 1.500 &  &  &  & 0.30 &   11.0 & 2.75 \\
18  & 0.50 & 0.07 & 1.46 & 1.45 & 1.622 & 5.252 - 1.877 & 7.111 - 5.573 & 0.497 - 0.499 &  0.17 &   1.0 & 10.9 \\
\cline{2-12}
    & 1.32 & 0.93 & 1.46 & 1.45 & 1.450 & & & & 1.09 &  20.0 & 5.42 \\
    & 1.32 & 0.93 & 1.46 & 1.45 & 1.500 &  &  &  & 0.32 & 17.0 & 1.85 \\
    & 1.32 & 0.93 & 1.46 & 1.45 & 1.622 & 6.176 - 2.609 & 7.278 - 5.915 &  0.497 - 0.499 &  0.27 &  8.0 & 3.49 \\
\hline
    & 1.32 & 0.70 & 1.58 & 1.60 & 1.600 &  &  &  & 1.14 &  22.0 & 5.25 \\
    & 1.32 & 0.70 & 1.58 & 1.60 & 1.650 &  &  &  &  0.48 & 20.0 & 2.35 \\
22  & 1.32 & 0.70 & 1.58 & 1.60 & 1.820 & 5.704 - 2.651 & 7.199 - 5.957 & 0.498 - 0.499 &  0.46 & 9.0 & 4.85 \\
\cline{2-12}
    & 3.50 & 2.99 & 1.58 & 1.60 & 1.600 & &  &  & 2.35  & 32.0 & 7.29 \\
    & 3.50 & 2.99 & 1.58 & 1.60 & 1.650 &  &  &  & 1.14 & 32.0 & 3.60 \\
    & 3.50 & 2.99 & 1.58 & 1.60 & 1.820 & 6.950 - 2.979 & 7.440 - 5.940 & 0.498 - 0.499  & 1.00 & 20.0 & 4.90 \\
\hline
     & 3.50 & 2.79 & 1.61 & 1.60 & 1.600 &  &  &  &  10.6 & 34.0 & 31.0\\
25   & 3.50 & 2.79 & 1.61 & 1.60 & 1.650 & &  &  & 9.99 & 34.0 & 29.4\\
     & 3.50 & 2.79 & 1.61 & 1.60 & 1.814 & 6.675 - 1.250 & 7.360 - 4.813  & 0.498 - 0.499  & 9.88 & 25.0 & 39.0\\
\end{tabular}
\end{center}
\end{table*}

\begin{figure}[H]
\centering
\epsscale{1.1}
\plotone{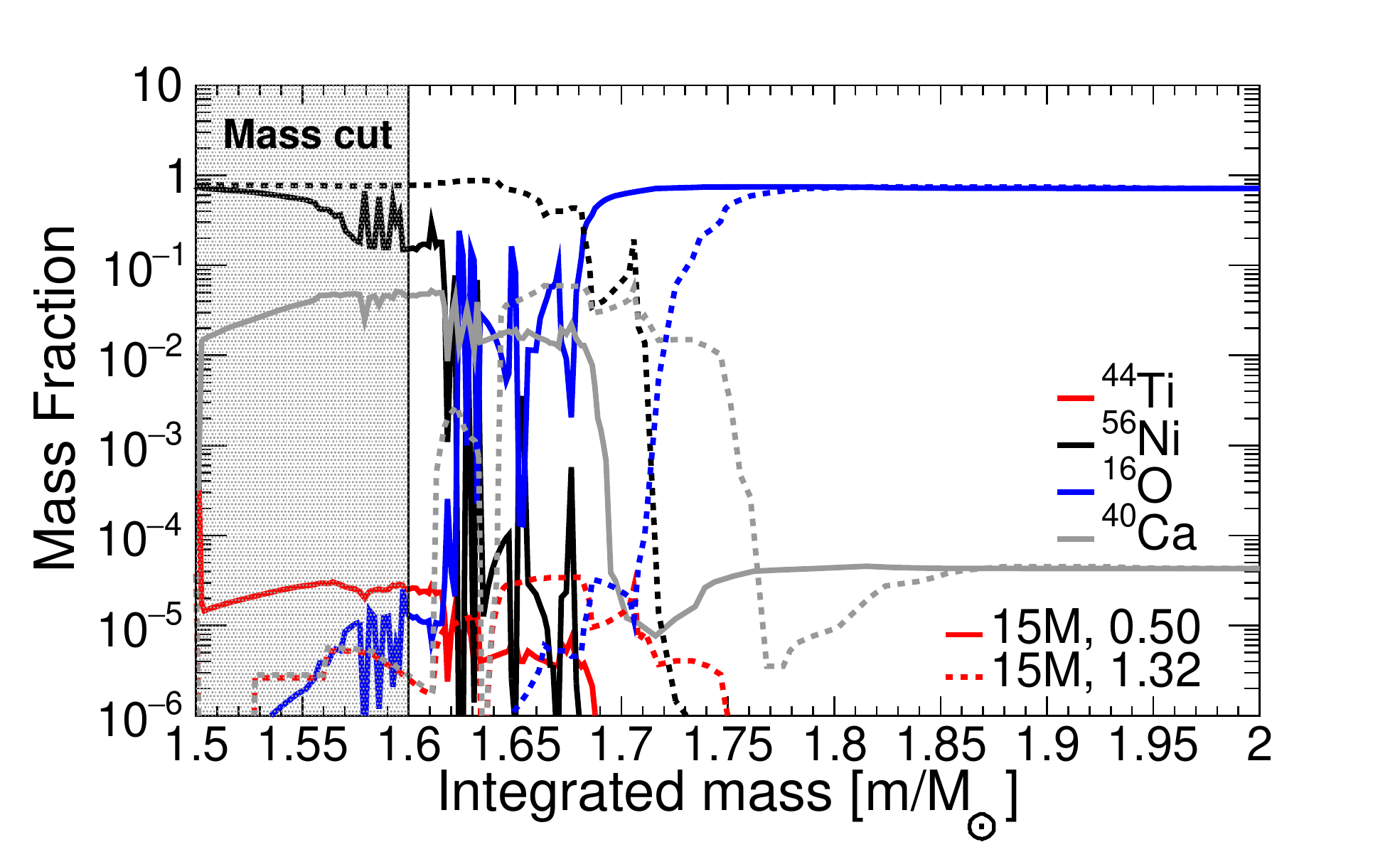}
\caption{Mass fraction vs integrated mass for various isotopes 5 seconds after the deposition of $E_{\rm inj}$ for calculations with $M_{\rm prog}=15~M_{\odot}$. $E_{\rm inj}$ is indicated by the line type: 0.50 foe (solid) and 1.32 foe (dotted).}
\label{fig:abund_15M}
\end{figure}

\section{Yields of ${^{44}}{\rm Ti}$ and ${^{56}}{\rm Ni}$ from Baseline Calculations} \label{sec:nuc_yield}

Prior to discussing the impact of varied nuclear reaction rates on $^{44}{\rm Ti}$ and $^{56}{\rm Ni}$ production, we first present results from our calculations using the baseline reaction rate library. The yields presented in Table~\ref{tab:abund} were determined by integrating the total mass of $^{44}{\rm Ti}$ or $^{56}{\rm Ni}$ outward from a lower-bound in mass known as the mass-cut $M_{\rm cut}$~\citep{Diehl_1998}. To help assess the impact of this arbitrary boundary on our results, we explored three choices of $M_{\rm cut}$: $I_{\rm b}$, $I_{\rm b}+\Delta M_{\rm shell}$, and $M_{4}$. The latter is defined as the mass zone where entropy per nucleon $s=4$, beyond which higher-dimensional models have found most material avoids fallback onto the proto-neutron star~\citep{2016ApJ...818..124E}. In order to compare with post-processing calculations, we determine $X(^{44}{\rm Ti})$ and $X(^{56}{\rm Ni})$ over the region between $M_{\rm cut}$ and the mass coordinate where there is no longer oxygen-burning, determining the mass-fraction per zone and taking an average weighted by the mass of each zone.

Figure \ref{fig:abund_15M} shows selected mass fractions following shock-driven nucleosynthesis for example calculations, with one value of $M_{\rm cut}$ shown for context. Clearly the choice of $M_{\rm cut}$ and $E_{\rm inj}$ will significantly affect yields of $^{44}{\rm Ti}$ and $^{56}{\rm Ni}$, highlighting the need to explore nuclear sensitivities for multiple scenarios.

\begin{figure}
\centering
\epsscale{1.2}
\plotone{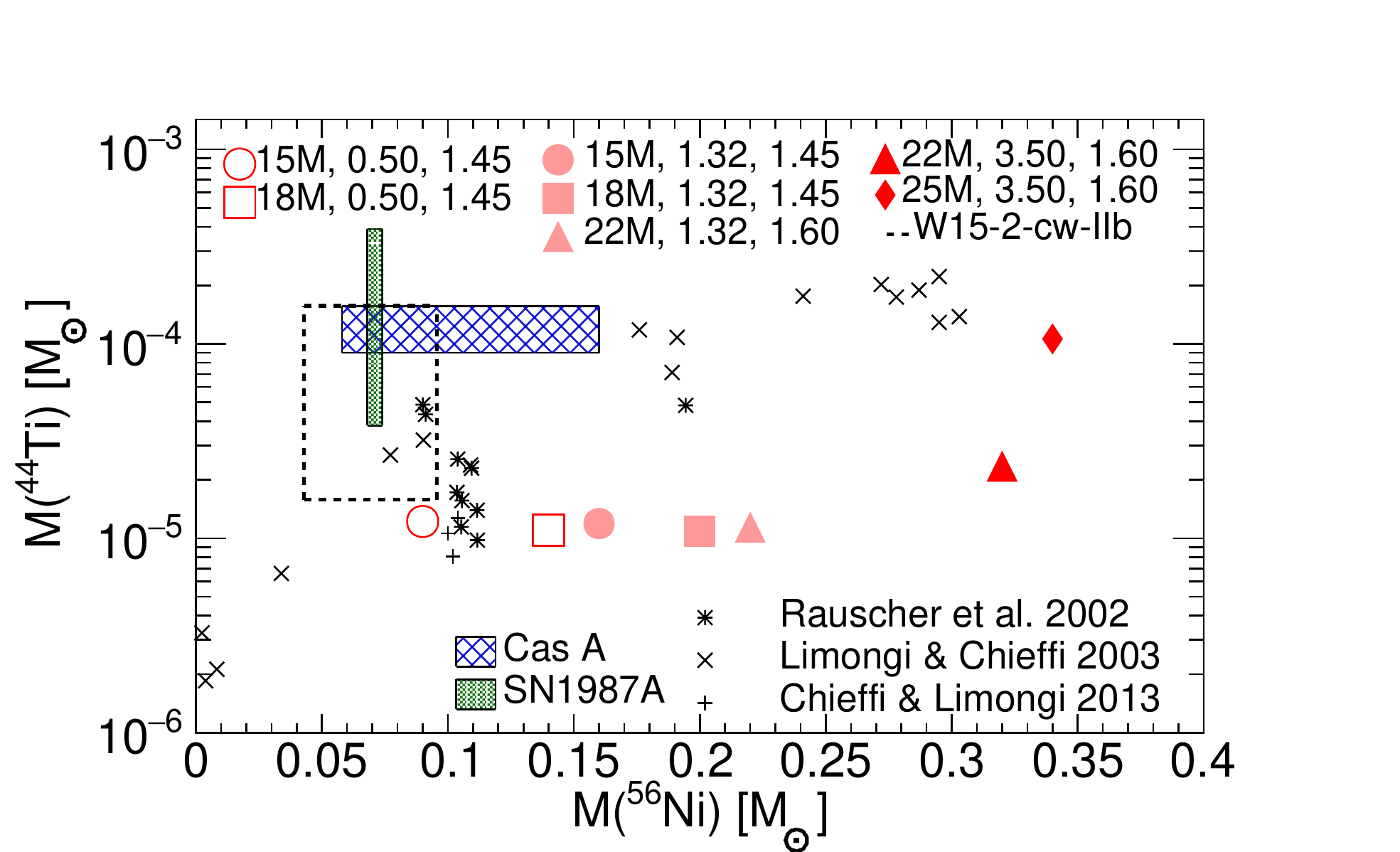}
\caption{Total mass of $^{44}{\rm Ti}$ and $^{56}{\rm Ni}$ outside of $M_{\rm cut}=I_{\rm b}$ from our calculations, where the legend indicates $M_{\rm prog}$, $E_{\rm inj}$, $M_{\rm cut}$. Results from previous observational and modeling work is shown for context. See text for details.}
\label{fig:ti44_ni56}
\end{figure}

Figure~\ref{fig:ti44_ni56} compares yields of $^{44}{\rm Ti}$ and $^{56}{\rm Ni}$ from our baseline calculations using various combinations of $M_{\rm prog}$, $E_{\rm inj}$, and $M_{\rm cut}$ to yields from model calculations of earlier studies, as well as inferred yields from astronomical observations. The boundary for observations of 1987A is based on the results of \citet{Boggs670} ($M({^{44}}{\rm Ti}) = (1.5\pm0.3)\times10^{-4}~M_{\odot}$), \citet{2011A&A...530A..45J} ($M({^{44}}{\rm Ti}) = (1.5\pm0.5)\times10^{-4}~M_{\odot}$), ~\citet{2012Natur.490..373G} ($M({^{44}}{\rm Ti}) = (3.1\pm0.8)\times10^{-4}~M_{\odot}$), and \citet{2014ApJ...792...10S} ($M({^{44}}{\rm Ti}) = (0.55\pm0.17)\times10^{-4}~M_{\odot}$, $M({^{56}}{\rm Ni}) = (7.1\pm0.3)\times10^{-2}~M_{\odot}$). The boundary for observations of CasA is from \citet{2016ApJ...825..102W} ($M({^{44}}{\rm Ti}) = (1.3\pm0.4)\times10^{-4}~M_{\odot}$), \citet{2015A&A...579A.124S} ($M({^{44}}{\rm Ti}) = (1.37\pm0.19)\times10^{-4}~M_{\odot}$), ~\citet{Grefenstette2014AsymmetriesIC} ($M({^{44}}{\rm Ti}) = (1.25\pm0.3)\times10^{-4}~M_{\odot}$), and~\citet{Eriksen_2009} ($M({^{56}}{\rm Ni}) = 5.8-16.0\times10^{-2}~M_{\odot}$).  The boundary labeled W15-2-cw-llb is from the 3D model calculation of \citet{2017ApJ...842...13W} that resembled CasA. We note that comparisons to CasA may not be particularly relevant for this work, as this supernova is suspected to have originated from a progenitor that lost much of its hydrogen envelope~\citep{Koo2020}. Results from 1D model calculations of \citet{Rauscher_2002}, ~\citet{2003ApJ...592..404L}, and~\citet{Chieffi_2013} are also shown for $M_{\rm prog} = 15-25~M_{\odot}$. While our $^{56}{\rm Ni}$ yields are similar to these previous 1D calculations, our $^{44}{\rm Ti}$ yields are generally lower, especially for our 18 and 22~$M_{\odot}$ progenitors. This is likely due to the different explosion mechanism (the other works do not employ a thermal bomb) and our choice of $E_{\rm inj}$ and $t_{\rm inj}$~\citep{2007ApJ...664.1033Y,2012ApJ...749...91F,Fryer_2018}, but a more comprehensive cross-model comparison, which is beyond the scope of the present work, would be necessary to comment further.

To provide further context for comparisons to earlier 1D modeling work, we calculate the explosion energy $E_{\rm exp}$ as described by \citet{1991ApJ...370..630A} and restated in Equations~\ref{eqn:e_exp1} and \ref{eqn:e_exp2}:
\begin{equation} 
\label{eqn:e_exp1}
E_{\rm exp}= E_{\rm inj} + E_{\rm b} + \Delta E_{\rm n},
\end{equation}
where
\begin{equation} 
\label{eqn:e_exp2}
E_{\rm b}= E_{\rm k} - E_{\rm Gv} + E_{\rm int}.
\end{equation}
Here $E_{\rm k}$ is the total kinetic energy, $E_{\rm Gv}$ is the total gravitational energy, $E_{\rm int}$ is the total internal energy, $E_{\rm b}$ is the total binding energy, and $\Delta E_{\rm n}$ is the nuclear binding energy released by burning. All the energy values are calculated from $I_{\rm b}$ to the surface of a star. Compared to previous works, e.g. \citet{1991ApJ...370..630A,Rauscher_2002,2003ApJ...592..404L}, our calculations with the lower of two $E_{\rm inj}$ for a given $M_{\rm prog}$ result in relatively low $E_{\rm exp}$, though the high $E_{\rm inj}$ $M_{\rm prog}=22~M_{\odot}$ and $25~M_{\odot}$ cases result in relatively high $E_{\rm exp}$.

\section{reaction rate variations} \label{sec:rr_var}

For each combination of $M_{\rm prog}$ and $E_{\rm inj}$ listed in Table~\ref{tab:abund}, we performed reaction rate variations and assessed the change in $^{44}{\rm Ti}$ and $^{56}{\rm Ni}$ yields relative to the baseline calculation. Given the computational expense, we chose to vary a single reaction rate at a time, as opposed to varying several rates at once in a Monte Carlo fashion \citep[e.g.][]{Blis20}. We limited our investigation to the 49 reaction rates identified as high impact by \citet{Magkotsios_2010}, excluding the weak rates identified in that work, along with two additional reaction rates chosen to test the completeness of the 49-rate list. The reaction list, given in Table~\ref{tab:table_rr}, includes the additional two reactions $^{27}{\rm Al}(\alpha,n)^{30}{\rm P}$ and $^{39}{\rm K}(p,\alpha)^{36}{\rm Ar}$ based on suspected impact due to the reaction network flow (See Figure~\ref{fig:net_flow}).

Similar past studies have often adopted reaction rate variation factors $\times100$ and $/100$~\citep[e.g.][]{The_1998,Magkotsios_2010}, since these studies aimed to find all plausible sensitivities. However, concerted efforts from the nuclear physics community have reduced the uncertainties for several reaction rates well below two orders of magnitude uncertainty. Furthermore, decades of comparison between measured data and Hauser-Feshbach calculations have shown that nuclear reactions proceeding through compound nuclei at relatively high nuclear level densities have theoretical predictions generally within an order of magnitude of measurements (see e.g. \citet{Mohr15}). As such, we aimed to adopt more realistic rate variation factors, taking into account existing experimental and theoretical data.

Reaction rate variation factors used here are given in Table~\ref{tab:table_rr}. When experimental data were available over the astrophysically relevant energies or a reaction rate evaluation existed, the uncertainty from that work was used as the rate variation factor. In the absence of such information, the rate was either assigned an uncertainty factor of 10 or 100, depending on whether or not the Hauser-Feshbach formalism was thought to be valid, following the approach of \citet{0004-637X-830-2-55}.

For Hauser-Feshbach validity, we adopt the heuristic that more than ten nuclear levels must exist within an MeV of the excitation energy populated in the compound nucleus for a particular reaction~\citep{Wago69,Raus97}.
To determine the relevant excitation energy, we use the Gamow window approximation for the center of mass reaction energy of astrophysical interest \citep{cauldrons}:
  \begin{equation}
  E_{\rm G}= 0.1220\left(Z_{1}^{2}Z_{2}^{2}\frac{A_{1}A_{2}}{(A_{1} + A_{2})}T_{9}^{2}\right)^{1/3}~{\rm MeV},
  \end{equation}
where $Z_{i}$ are the proton numbers of the two reactants, $A_{i}$ are their mass numbers, $T_{9}$ is the environment temperature in gigakelvin, and $k_{\rm B}$ is the Boltzmann constant. The $1/e$ width of the window about $E_{\rm G}$ is
\begin{equation} 
\label{eqn:gam1}
\Delta_{G} = 4 \sqrt{\left(\frac{E_{\rm G}\hspace{1mm}k_{\rm B}\hspace{1mm}T_{9}}{3}\right)}~{\rm MeV}.
\end{equation}
While more accurate methods exist to determine the excitation energy window of astrophysical interest~\citep{Newt07,Raus10}, we chose the more approximate Gamow window method given the uncertainty of nuclear level structure for the nuclides involved and the choice of a single temperature for which to calculate the energies of astrophysical interest.

We calculated the Gamow window at $T_{9}$ = 4 as this is where the calculations of \citet{Magkotsios_2010} fall out of QSE. We determined whether at the bottom of the window ($E_{\rm G}-\Delta_{\rm G}/2$) more than 10 nuclear levels per MeV of excitation were present, based on the microscopic nuclear level densities calculated by \citet{Gori08}. For high level-density cases, we consider a reaction to be in the statistical regime and assign an uncertainty factor of 10, based on the typical spread in predictions from Hauser-Feshbach calculations~\citep[e.g.][]{Pere16}. For low level-density cases, we consider a reaction to be in the resonant regime and assign an uncertainty factor of 100, based on the significant rate modification that can occur due addition/removal of a particular resonance or change in a key resonance's properties~\citep[e.g.][]{Cava15}.

  \startlongtable 
\begin{deluxetable*}{ccc||ccc||ccc}
\tablecaption{Reaction rate variation factors (RR). The rate numbers are used as identifiers in Figures (\ref{fig:sen_ti44}), (\ref{fig:sen_ni56}), and (\ref{fig:sen_ratio}). \label{tab:table_rr}}

\startdata
 & $(\alpha, n)$, $(n, \alpha)$ &  &  & $(\alpha, p)$, $(p,\alpha)$ &  &  & $(p, \gamma)$ & \\ 
\hline
Reactions &RR($\uparrow$)($\downarrow$) & Reac \# & Reactions & RR($\uparrow$)($\downarrow$) &Reac \# &  Reactions & RR($\uparrow$)($\downarrow$) & Reac \#  \\
\hline
 ${^{10}} {\rm B}(\alpha,n){^{13}}{\rm N}$ & 20\% (1) & 1 & ${^{44}}{\rm Ti}(\alpha,p){^{47}}{\rm V}$ & 44\% (10)\tablenotemark{a} & 7 & ${^{45}}{\rm V}(p,\gamma){^{46}}{\rm Cr}$ & 10  & 33\\
${^{11}}{\rm B}(\alpha,n){^{14}}{\rm N}$ & 10\% (2) & 2 & ${^{40}}{\rm Ca}(\alpha,p){^{43}}{\rm Sc}$ & 15\% (11) & 8 & $^{41}{\rm Sc}(p,\gamma)^{42}{\rm Ti}$ & 100 & 34\\
 ${^{23}}{\rm Mg}(n,\alpha){^{20}}{\rm Ne}$ & 100 & 3 & ${^{17}}{\rm F}(\alpha,p){^{20}}{\rm Ne}$ & 100 & 9 & ${^{43}}{\rm Sc}(p,\gamma){^{44}}{\rm Ti}$ & 10 & 35\\
${^{9}}{\rm Be}(\alpha,n){^{12}}{\rm C}$ & 10\% (3) & 4 & ${^{21}}{\rm Na}(\alpha,p){^{24}}{\rm Mg}$ & 100 & 10 & ${^{44}}{\rm Ti}(p,\gamma){^{45}}{\rm V}$ & 100 & 36 \\
${^{42}}{\rm Ca}(\alpha,n){^{45}}{\rm Ti}$ & 21\% (4) & 5 & ${^{27}}{\rm Al}(\alpha,p){^{30}}{\rm Si}$ & 100 & 11 & ${^{57}}{\rm Ni}(p,\gamma){^{58}}{\rm Cu}$ & 10 & 37 \\
${^{34}}{\rm S}(\alpha,n)^{37}{\rm Ar}$ & 16\% (5) & 6 &${^{55}}{\rm Co}(\alpha,p){^{58}}{\rm Ni}$ & 10 & 12 & ${^{40}}{\rm Ca}(p,\gamma){^{41}}{\rm Sc}$ & 100 & 38\\
${^{27}}{\rm Al}(\alpha,n){^{30}}{\rm P}$ & 10 & 50 & ${^{48}}{\rm Cr}(\alpha,p){^{51}}{\rm Mn}$ & 100 & 13 &                                        ${^{44}}{\rm V}(p,\gamma){^{45}}{\rm Cr}$ & 100 & 39 \\          
\cline {1-3} \cline {1-3}
& $(\alpha, \gamma)$  & &  ${^{52}}{\rm Fe}(\alpha,p){^{55}}{\rm Co}$ & 100  & 14  &  ${^{43}}{\rm Ti}(p,\gamma){^{44}}{\rm V}$ & 10 & 40\\
\cline {1-3} 
Reactions  & RR($\uparrow$)($\downarrow$) & Reac \# &   ${^{54}}{\rm Fe}(\alpha,p){^{57}}{\rm Co}$ & 8\% (12) & 15  & ${^{42}}{\rm Sc}(p,\gamma){^{43}}{\rm Ti}$ & 10 & 41\\
\cline {1-3} 
${^{40}}{\rm Ca}(\alpha,\gamma){^{44}}{\rm Ti}$ & 25\% (6) & 24 &  ${^{56}}{\rm Ni}(\alpha,p){^{59}}{\rm Cu}$ & 10   & 16 & ${^{57}}{\rm Cu}(p,\gamma){^{58}}{\rm Zn}$ & 100 & 42 \\
${^{12}}{\rm C}(\alpha,\gamma){^{16}}{\rm O}$ & 20\% (7)(8) & 25 &   ${^{6}}{\rm Li}(\alpha,p){^{9}}{\rm Be}$ & 100 & 17 &  ${^{20}}{\rm Ne}(p,\gamma){^{21}}{\rm Na}$ 
& 12\% (15)\tablenotemark{b}   & 43\\
${^{7}}{\rm Be}(\alpha,\gamma){^{11}}{\rm C}$ & 100 & 26 &   ${^{13}}{\rm N}(\alpha,p){^{16}}{\rm O}$ & 100  & 18       & ${^{47}}{\rm V}(p,\gamma){^{48}}{\rm Cr}$ & 10 & 44\\
${^{24}}{\rm Mg}(\alpha,\gamma){^{28}}{\rm Si}$ & 10\tablenotemark{c} & 27 &   ${^{42}}{\rm Ca}(\alpha,p){^{45}}{\rm Sc}$ & 21\% (13)   & 19     & ${^{42}}{\rm Ca}(p,\gamma){^{43}}{\rm Sc}$ & 21\% (16) & 45\\
${^{42}}{\rm Ca}(\alpha,\gamma){^{46}}{\rm Ti}$ & 15\% (9) & 28 &  ${^{43}}{\rm Sc}(\alpha,p){^{46}}{\rm Ti}$ & 10  & 20     & ${^{39}}{\rm K}(p,\gamma){^{40}}{\rm Ca}$ 
& 100\tablenotemark{d}  & 46\\
\cline {1-3} \cline {1-3}
& $(p, n)$, $(n, p)$  &   &  ${^{58}}{\rm Ni}(\alpha,p){^{61}}{\rm Cu}$ & 25\% (14) & 21 & ${^{57}}{\rm Co}(p,\gamma){^{58}}{\rm Ni}$ & 10 & 47\\
\cline {1-3} 
Reactions  & RR($\uparrow$)($\downarrow$) & Reac \# &  ${^{38}}{\rm Ca}(\alpha,p){^{41}}{\rm Sc}$ & 100  & 22      &${^{54}}{\rm Fe}(p,\gamma){^{55}}{\rm Co}$ & 12\% (17) & 48 \\
\cline {1-3} 
${^{57}}{\rm Ni}(n,p){^{57}}{\rm Co}$ & 100 & 29 &   ${^{34}}{\rm Ar}(\alpha,p){^{37}}{\rm K}$ & 100 &  23 &  ${^{52}}{\rm Mn}(p,\gamma){^{53}}{\rm Fe}$ & 10 & 49 \\
${^{56}}{\rm Co}(p,n){^{56}}{\rm Ni}$ & 100 & 30 &${^{39}}{\rm K}(p, \alpha){^{36}}{\rm Ar}$ & 100 & 51 &\\
${^{27}}{\rm Si}(n,p){^{27}}{\rm Al}$ & 100 & 31 &   &  &\\
$^{11}{\rm C}(n,p)^{11}{\rm B}$ & 100 & 32 & & & & & & \\
\enddata 
\tablerefs{(1) \cite{PhysRev.114.571}, (2) \cite {PhysRevC.43.883}, (3) \cite{PhysRevC.49.1205}, (4) \cite{7a0fce1bb6fb4f259feff635b3352b0e}, (5) \cite{SCOTT1993363}, (6) \cite{PhysRevC.85.045810}, (7) ~\cite{PhysRevC.86.015805}, (8) ~\cite{2017RvMP...89c5007D}, (9) ~\cite{MITCHELL1985487}, (10) ~\cite{PhysRevLett.84.1651}, (11) ~\cite{1974ApJ...188..131H}, (12) ~\cite{TIMS1991479}, (13) ~\cite{0305-4616-9-1-013}, (14)~\cite{VLIEKS1974492}, (15) ~\cite{ROLFS1975460}, (16) ~\cite{VLIEKS1978506}, (17) ~\cite{KENNETT1981233}}    
\tablenotetext{a}{Subsequent evaluations~\citep{Hoff10,Chip20} suggest that \citet{PhysRevLett.84.1651} underestimate the rate uncertainty and that a factor of 3 or more would be more appropriate.}
\tablenotetext{b}{\cite{PhysRevC.97.065802} have since updated the rate, but the rate and uncertainty are similar in our temperature range of interest.}
\tablenotetext{c}{The evaluation of \citet{Adsl20} suggests a much smaller factor may be appropriate; however, they focused on lower temperatures than are of interest for this work.}
\tablenotetext{d}{Our calculations were performed prior to the rate evaluation of \citet{2018PhRvC..98b5802L}, who suggest $\sim\times10$ is likely a more appropriate factor for the temperature range of interest. However, we note that experimental coverage is incomplete for the relevant center of mass energies.}
\end{deluxetable*}

\section{Sensitivity study results} \label{sec:sen_study}

\begin{figure*}[ht]
\centering
\plotone{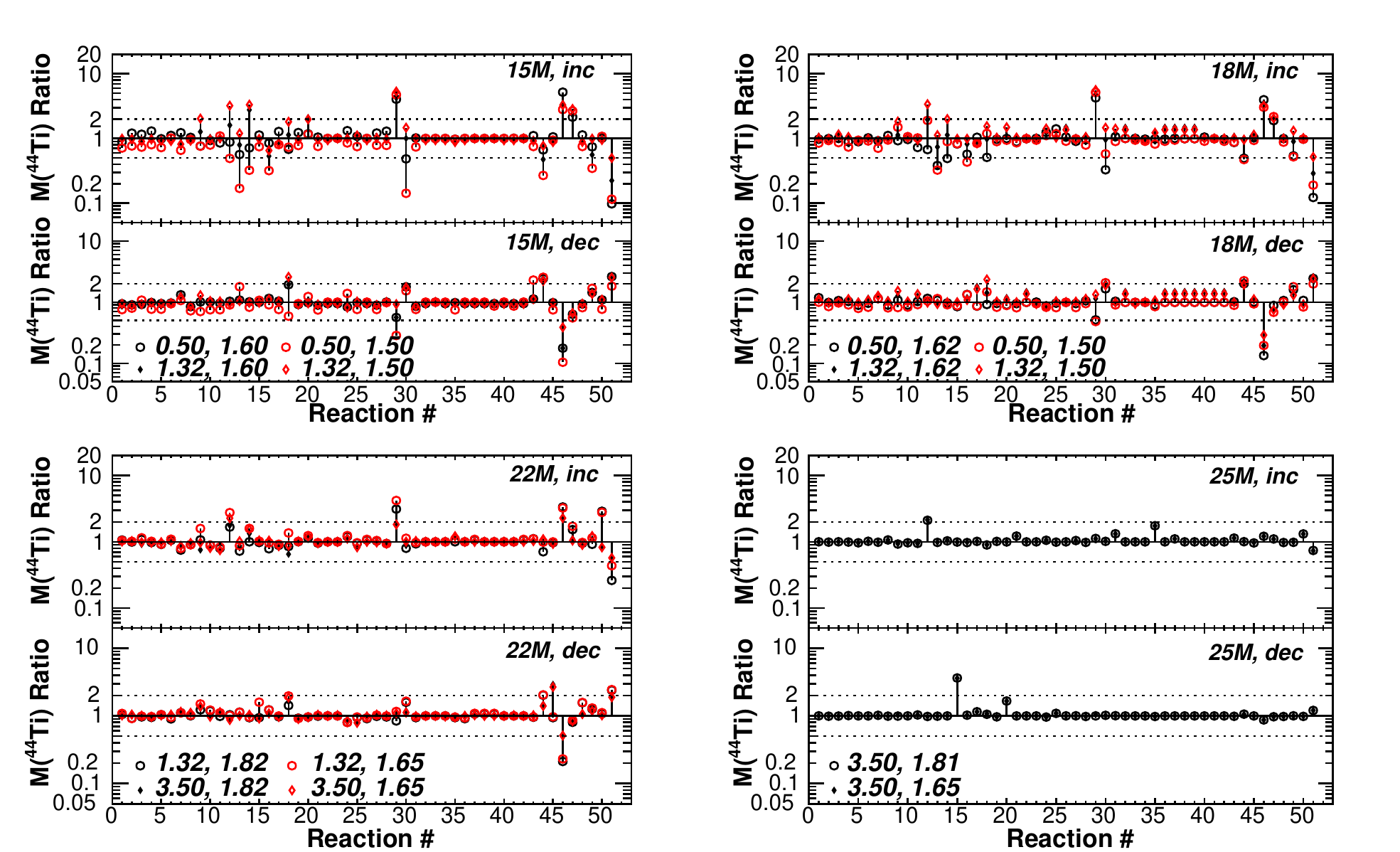}
\caption{Ratio of $M(^{44}\rm{Ti})$ to the baseline calculation results of Table~\ref{tab:abund} when varying each reaction of Table~\ref{tab:table_rr} by its uncertainty factor upward (upper panel) or downward (lower panel). The legends indicate $E_{\rm inj}$, $M_{\rm cut}$. The dashed lines demarcate factor of two impacts.}
\label{fig:sen_ti44}
\end{figure*}

\begin{figure*}[ht]
\centering
\plotone{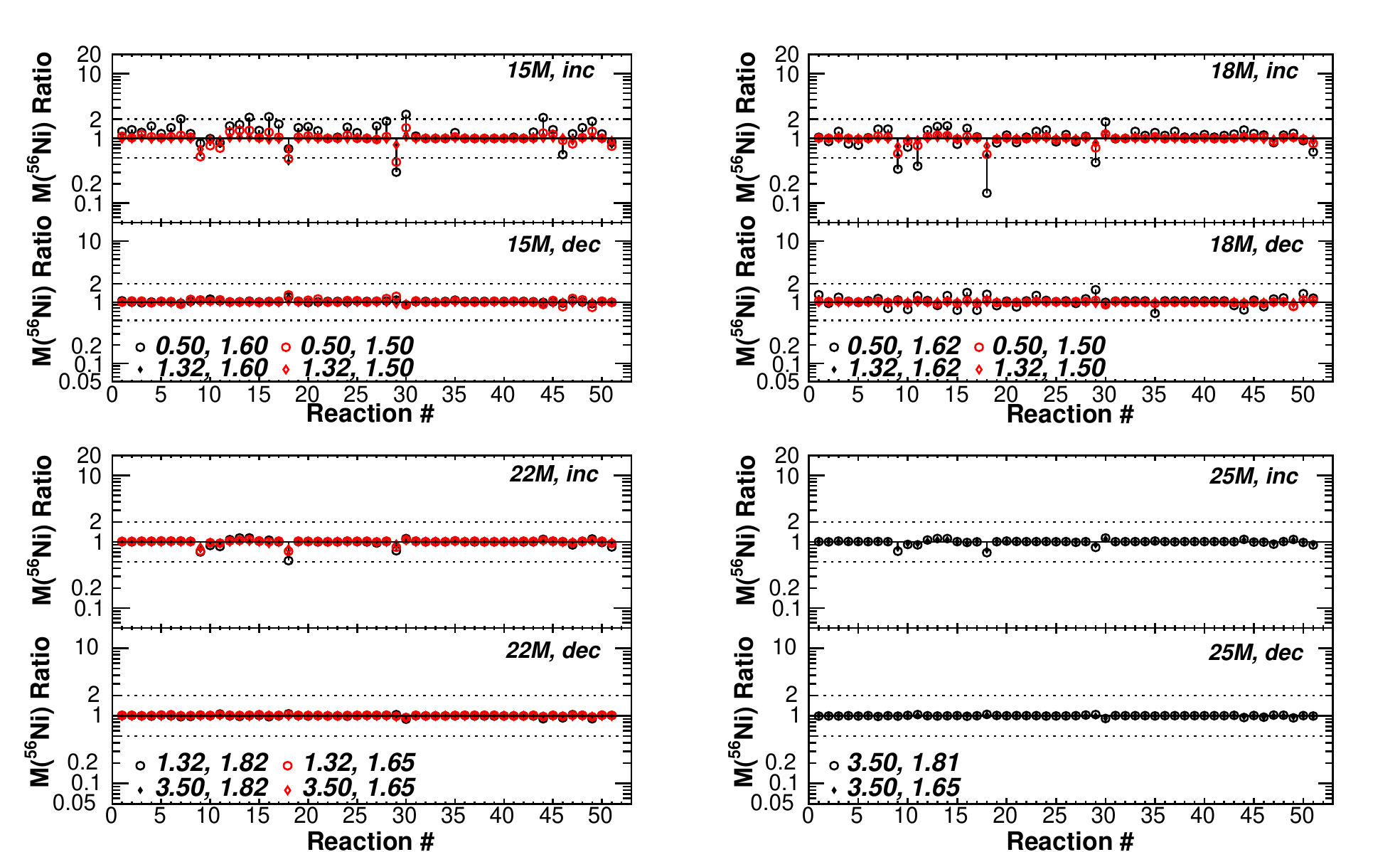}
\caption{Same as Figure~\ref{fig:sen_ti44} but for $M(^{56}{\rm Ni})$.}
\label{fig:sen_ni56}
\end{figure*}

\begin{figure*}[ht]
\centering
\plotone{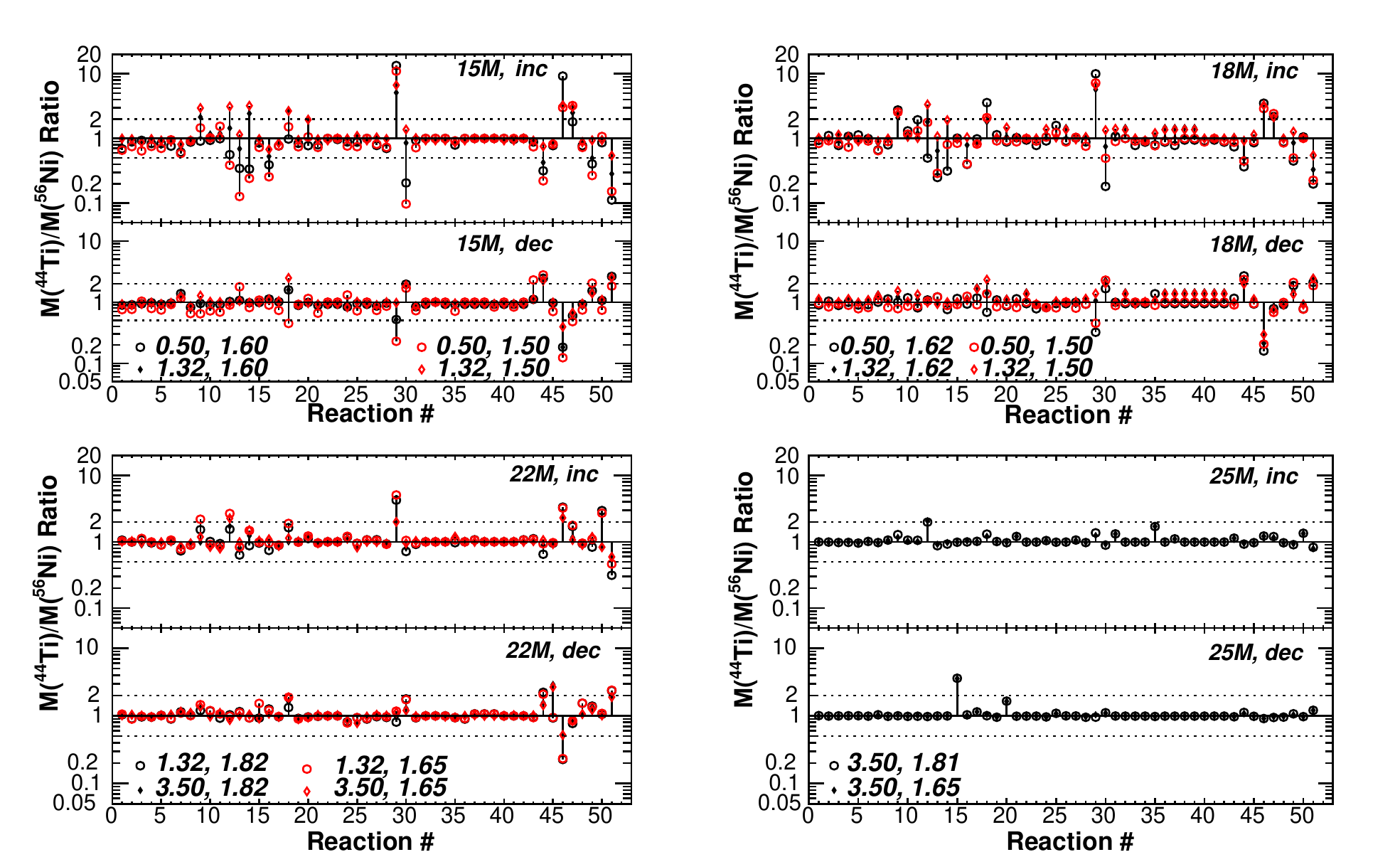}
\caption{Same as Figure~\ref{fig:sen_ti44} but for $M(^{44}{\rm Ti})/M(^{56}{\rm Ni})$.}
\label{fig:sen_ratio}
\end{figure*}

Figures \ref{fig:sen_ti44}, \ref{fig:sen_ni56}, and \ref{fig:sen_ratio} show the sensitivity study results for ${^{44}}$Ti, ${^{56}}$Ni, and ${^{44}}\rm{Ti}/{^{56}}\rm{Ni}$, respectively, for all combinations of $M_{\rm prog}$, $E_{\rm inj}$, and $M_{\rm cut}$ modeled here, where the reaction numbers correspond to the designations in Table~\ref{tab:table_rr} and the reported ratios are to the baseline results of Table~\ref{tab:abund}. In each plot the upper panel corresponds to a reaction rate increase, while the bottom panel corresponds to a reaction rate decrease. We show the sensitivities for various combinations of $E_{\rm inj}$ and $M_{\rm cut}$ to give a sense of the robustness of our results to changes in model assumptions. However, for the ensuing discussion we limit ourselves to the results for $M_{\rm cut}=M_{4}$.

In the following we separately discuss reactions impacting $^{44}{\rm Ti}$, $^{56}{\rm Ni}$, and their ratio. Motivated by typical observational uncertainties (discussed in Section~\ref{sec:nuc_yield}), we only discuss cases (listed in Table~\ref{tab:sen_val}) that affect these yields by a factor of two or more for at least one set of model conditions, as we deem this to be significant for model-observation comparisons. Of course as observations improve, it is likely that even smaller sensitivities will be of interest in the future. Note that in general we see fewer reaction rate variations having a significant impact for larger $M_{\rm prog}$ calculations. This is because, for those cases, much of the $^{44}{\rm Ti}$ and $^{56}{\rm Ni}$ beyond $M_{\rm cut}$ was synthesized during stellar evolution at radii not significantly impacted by the outgoing shock~\citep{Chieffi_2017}.

The explanations we provide for the sensitivities we observe are based on analyzing the flow in the nuclear reaction network (see Figure~\ref{fig:net_flow}), along with guidance from the discussions in \citet{The_1998} and \citet{Magkotsios_2010}.
The flow for converting isotope $i$ to isotope $j$ over a timestep is
$\mathcal{F}_{ij}=\int\left(\dot{Y}_{i\rightarrow
j}-\dot{Y}_{j\rightarrow i}\right)dt$, where  $\dot{Y}_{i\rightarrow j}$ is the rate that the
abundance of isotope $i$ is depleted by reactions converting $i$ to
$j$.
As shown in Figure~\ref{fig:temp_dens}, the relevant conditions for this work are primarily those known as $\alpha$-rich and silicon-rich freeze-out. For silicon-rich conditions (Region A), nucleosynthesis yields are mostly characterized by nuclear masses and the initial abundances before the explosion, with some flow between many small QSE clusters. 
For $\alpha$-rich freeze-out (Region D), much larger QSE clusters are formed initially and a significant excess of $\alpha$-particles are available. In that case nucleosynthesis is characterized by the flow between these large QSE clusters and the fusion of $\alpha$-particles into heavier nuclides with $Z=2n, A=4n$, where $n$ is an integer. Therefore, our reaction rate sensitivities fall into three categories: those influencing the helium burning, those near boundaries of the large QSE clusters that exist in $\alpha$-rich freeze-out at early times, and those near $^{44}{\rm Ti}$ that can influence nucleosynthesis as that QSE cluster dissolves into many smaller QSE clusters.

\subsection{Reactions impacting ${^{44}}$\rm{Ti}} \label{sec:ti44}

Here we discuss reactions listed in Table~\ref{tab:sen_val} that significantly impact $M(^{44}{\rm Ti})$, working from light to heavy nuclides.

${^{13}}{\rm N}(\alpha,p){^{16}}{\rm O}$ and $^{27}{\rm Al}(\alpha,n)^{30}{\rm P}$ each influence $M(^{44}{\rm Ti})$ for a subset of conditions, where a rate increase of the former reduces $^{44}{\rm Ti}$ yields, while an in increase of the latter increases $M(^{44}{\rm Ti})$. Each is a prominent helium-burning reaction in our network flow. However, the lack of consistency in influence across $M_{\rm prog}$ and $E_{\rm inj}$ demonstrates a clear dependence on the adopted model conditions.

$^{39}{\rm K}(p,\gamma)^{40}{\rm Ca}$ and $^{39}{\rm K}(p,\alpha)^{36}{\rm Ar}$ significantly impact $M(^{44}{\rm Ti})$ across almost all sets of model conditions. This is due to the importance of the reaction sequence $^{39}{\rm K}(p,\gamma)$ $^{40}{\rm Ca}(\alpha,\gamma)^{44}{\rm Ti}$. Increasing $^{39}{\rm K}(p,\gamma)^{40}{\rm Ca}$ clearly enhances $M(^{44}{\rm Ti})$, whereas $^{39}{\rm K}(p,\alpha)^{36}{\rm Ar}$ competes as an alternative reaction path, reducing $M(^{44}{\rm Ti})$. In each case, a reaction rate decrease has the opposite effect.

$^{42}{\rm Ca}(p,\gamma)^{43}{\rm Sc}$ is generally well-constrained enough so as to not impact $^{44}{\rm Ti}$ nucleosynthesis. However, for the $22~M_{\odot}$ high-$E_{\rm inj}$ case, decreasing this reaction rate significantly enhances $M(^{44}{\rm Ti})$. The connection of $^{42}{\rm Ca}$ to $^{44}{\rm Ti}$ along the reaction network is not as straight forward as one might assume, since $^{43}{\rm Sc}(\gamma,p)^{42}{\rm Ca}$ dwarfs the forward reaction at the relevant temperatures. The influence of decreasing the $^{42}{\rm Ca}(p,\gamma)^{43}{\rm Sc}$ reaction rate is to reduce competition with $^{42}{\rm Ca}(\alpha,n)^{45}{\rm Ti}$. The latter reaction enhances $M(^{44}{\rm Ti})$ since the abundances across titanium isotopes will be redistributed according to $(n,\gamma)-(\gamma,n)$ equilibrium until the environment has cooled and nucleosynthesis has essentially ceased.

\begin{figure}
\includegraphics[width=.37\textwidth, angle=270]{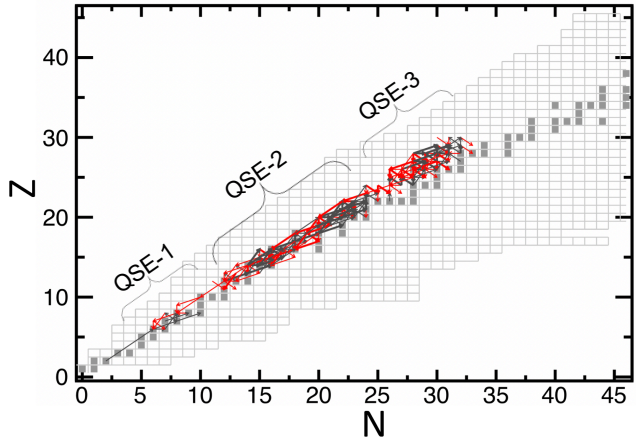}
\caption{Reaction network flow for $M_{\rm prog}=15~M_{\odot}$ with $E_{\rm inj}=1.32$~foe 0.06~seconds after the completion of energy deposition. Forward reaction rate flow is shown by black arrows, while the red arrows represent the reverse reaction flow. We refer to $(p,\gamma)$, $(p,n)$, $(\alpha,\gamma)$, $(\alpha,p)$, and $(\alpha,n)$ as ``forward". Three QSE regions that are present early in the calculation are identified in the reaction flow. Note that these begin as a single large QSE cluster and dissolve into many smaller QSE clusters.}
\label{fig:net_flow}
\end{figure}

An increase in the $^{43}{\rm Sc}(\alpha,p)^{46}{\rm Ti}$ rate
was found to enhance $M(^{44}{\rm Ti})$ only for the $15~M_{\odot}$ high-$E_{\rm inj}$ case, where the presumed connection to $^{44}{\rm Ti}$ is as described in the previous paragraph for $^{42}{\rm Ca}(\alpha,n)^{45}{\rm Ti}$. Following the reaction network flow of Figure~\ref{fig:net_flow} also reveals the connection $^{43}{\rm Sc}(\alpha,p)^{46}{\rm Ti}(p,\gamma)^{47}{\rm V}(p,\alpha)^{44}{\rm Ti}$.

$^{47}{\rm V}(p,\gamma)^{48}{\rm Cr}$ significantly impacts $M(^{44}{\rm Ti})$ for most of the model conditions explored. This reaction competes with $^{47}{\rm V}(p,\alpha)^{44}{\rm Ti}$, instead bridging the QSE leakage from QSE cluster 2 to QSE-3 (see Figure~\ref{fig:net_flow}). Therefore, an increase in the $^{47}{\rm V}(p,\gamma)^{48}{\rm Cr}$ rate decreases $M(^{44}{\rm Ti})$, while a reaction rate decrease has the opposite effect.

$^{48}{\rm Cr}(\alpha,p)^{51}{\rm Mn}$ can be a part of the bridge connecting QSE-2 to QSE-3 and therefore an increase in this reaction rate decreases $M(^{44}{\rm Ti})$ for low $M_{\rm prog}$ cases. However, when using the nominal $^{48}{\rm Cr}(\alpha,p)^{51}{\rm Mn}$ reaction rate, $^{48}{\rm Cr}(n,p)^{48}{\rm V}$ plays a more significant role bridging QSE-2 to QSE-3, and so a reaction rate decrease has little impact.

After the sequence $^{48}{\rm V}(p,
\gamma)^{49}{\rm Cr}(p,n)^{49}{\rm V}(p,\gamma)^{50}{\rm Cr}(p,\gamma)$ $^{51}{\rm Mn}(n,p)^{51}{\rm Cr}(p,\gamma)^{52}{\rm Mn}$, $^{52}{\rm Mn}(p,\gamma)^{53}{\rm Fe}$ is near the upper end of the bridge between QSE-2 and QSE-3. As such, increasing this reaction rate decreases $M(^{44}{\rm Ti})$, while decreasing the rate has the opposite effect, though our threshold for significance is only crossed for one set of model conditions.

The impact of $^{52}{\rm Fe}(\alpha,p)^{55}{\rm Co}$ significantly depends on the adopted model conditions, alternately enhancing or reducing $M(^{44}{\rm Ti})$ from a reaction rate increase. In Region A, where the majority of the low-$M_{\rm prog}$ low-$E_{\rm inj}$ mass zones are located, $^{52}{\rm Fe}$ is expected to be one of the primary freeze-out nuclides~\citep{Magkotsios_2010}, and so it is sensible that for these cases increasing the $^{52}{\rm Fe}(\alpha,p)^{55}{\rm Co}$ reaction rate will move the reaction network flow away from $^{44}{\rm Ti}$. However, for Region D, in which higher $T_{0}$ are experienced and $(\gamma,p)$ reactions will play a more prominent role, enhancing $^{52}{\rm Fe}(\alpha,p)^{55}{\rm Co}$ ultimately leads to more significant flow toward $^{44}{\rm Ti}$.
This case highlights the complexities of nuclear reaction network flows in high-temperature environments.

$^{54}{\rm Fe}(\alpha,p)^{57}{\rm Co}$ is quite well constrained experimentally and variations within the present uncertainty are largely inconsequential for $^{44}{\rm Ti}$ nucleosynthesis. However, a decrease in this reaction rate significantly enhances $M(^{44}{\rm Ti})$ for the highest $M_{\rm prog}$ calculation. This is likely due to the increased competition from the reaction sequence $^{54}{\rm Fe}(p,\alpha)^{51}{\rm Mn}(\gamma,p)^{50}{\rm Cr}(p,\alpha)^{47}{\rm V}(p,\alpha)^{44}{\rm Ti}$.

As with $^{52}{\rm Fe}(\alpha,p)^{55}{\rm Co}$, the influence of $^{55}{\rm Co}(\alpha,p)^{58}{\rm Ni}$ on $M(^{44}{\rm Ti})$ depends strongly on the model conditions, having the same relative influence as the former reaction for a reaction rate increase.

The majority of the remaining reactions on isotopes of cobalt and nickel influence $M(^{44}{\rm Ti})$ as might be expected, with reactions moving toward $^{44}{\rm Ti}$ enhancing its abundance and reactions moving away decreasing it. These include $^{56}{\rm Co}(p,n)^{56}{\rm Ni}$, $^{56}{\rm Ni}(\alpha,p)^{59}{\rm Cu}$, and $^{57}{\rm Ni}(n,p)^{57}{\rm Co}$. On the other hand, an increase of the $^{57}{\rm Co}(p,\gamma)^{58}{\rm Ni}$
reaction rate increases $M(^{44}{\rm Ti})$. This somewhat counterintuitive result is apparently due to the flow connecting $^{58}{\rm Ni}$ to $^{44}{\rm Ti}$ via $^{54}{\rm Fe}$. Specifically, the reaction sequence $^{58}{\rm Ni}(\gamma,n)^{57}{\rm Ni}(\gamma,p)^{56}{\rm Co}(\gamma,p)^{55}{\rm Fe}(\gamma,n)^{54}{\rm Fe}$ and onto $^{44}{\rm Ti}$ as described in the preceding paragraph discussing $^{54}{\rm Fe}(\alpha,p)^{57}{\rm Co}$.

\bgroup
\def\arraystretch{0.92}
\begin{longrotatetable}
 \centering
 \small
 \tabcaption{Ratios to the $M_{\rm cut}=M_{4}$ baseline calculation results of Table~\ref{tab:abund} for reaction rates significantly impacting  $M({^{44}}\rm{Ti})$, $M({^{56}}\rm{Ni})$, and/or $M({^{44}}\rm{Ti})/M({^{56}}\rm{Ni})$. Ratios greater than a factor of two increase or reduction are bolded for readability. The RR column lists the reaction rate multiplication factor. Sub-columns under each $M_{\rm prog}$ indicate $E_{\rm inj}$ in foe.}
\label{tab:sen_val}

\begin{tabular}{c|c||c|c|c|c|c|c|c||c|c|c|c|c|c|c||c|c|c|c|c|c|c}

&   &  \multicolumn{7}{c||}{$M({^{44}} \rm {Ti})$/$M(^{44}{\rm Ti})|_{\rm baseline}$} & \multicolumn{7}{c||}{$M({^{56}} \rm {Ni})$/$M(^{56}{\rm Ni})|_{\rm baseline}$} & \multicolumn{7}{c}{$[M({^{44}} \rm {Ti})$/$M(^{56}{\rm Ni})$]/[$M(^{44}{\rm Ti})$/$M(^{56}{\rm Ni})]|_{\rm baseline}$}    \\
\cline{3-23}
 Reactions   & RR &\multicolumn{2}{c|}{15~M$_{\odot}$}  & \multicolumn{2}{c|}{18~M$_{\odot}$}   & \multicolumn{2}{c|}{22~M$_{\odot}$}  & 25~M$_{\odot}$  &\multicolumn{2}{c|}{15~M$_{\odot}$}  & \multicolumn{2}{c|}{18~M$_{\odot}$}   & \multicolumn{2}{c|}{22~M$_{\odot}$}  & 25~M$_{\odot}$ &\multicolumn{2}{c|}{15~M$_{\odot}$}  & \multicolumn{2}{c|}{18~M$_{\odot}$}   & \multicolumn{2}{c|}{22~M$_{\odot}$}  & 25~M$_{\odot}$   \\
\cline{3-23}
    &  & 0.5 & 1.32 & 0.5 & 1.32  & 1.32 & 3.5 & 3.5 & 0.5 & 1.32 & 0.5 & 1.32  & 1.32 & 3.5 & 3.5 & 0.5 & 1.32 & 0.5 & 1.32  & 1.32 & 3.5 & 3.5    \\
\hline \hline
 ${^{17}}{\rm F}(\alpha,p){^{20}}{\rm Ne}$ & $\times$ 100 & 0.8 & 1.3 & 0.9 & 1.1  &  1.1 & 0.8 & 0.9 & 0.8 & 0.6 & \textbf{0.3} & \textbf{0.5} & 0.7 & 0.8 & 0.7 & 0.9 & \textbf{2.1} & \textbf{2.8} & \textbf{2.1} & 1.5 & 1.0 & 1.3 \\
    &  $\times$ 0.01 & 1.2 & 1.0 & 1.1 & 1.3 & 1.2 & 1.3 & 1.0 & 1.4 & 1.1 & 1.1 & 1.0 & 1.0 & 1.0 & 1.0 & 0.8 & 1.0 & 1.0 & 1.2 & 1.2 & 1.3 & 1 \\
\hline
${^{27}}{\rm Al}(\alpha,p){^{30}}{\rm Si}$ & $\times$ 100 & 0.8 & 1.0 & 0.7 & 1.0   & 0.8 & 0.8 & 0.9 & 0.9 & 0.8 & \textbf{0.4} & 0.8  & 0.8 & 0.9  & 0.9 & 1.0 & 1.2 & 1.9 & 1.2 & 0.9 & 0.9 & 1.0 \\
    &   $\times$ 0.01 & 1.2 & 1.0  & 1.0  & 1.2  & 1.0 & 1.1 & 1.0 & 1.6 & 1.0 & 1.3  & 1.1 & 1.1 & 1.0 & 1.0 & 0.8 & 0.9 & 0.9 & 1.1 & 0.9 & 1.0 & 1.0\\

\hline
   ${^{55}}{\rm Co}(\alpha,p){^{58}}{\rm Ni}$ & $\times$ 10 & 0.9 & 1.6 & 0.7 & \textbf{2.0}   & 1.7 & 1.8  & \textbf{2.1} & 1.6 & 1.1 & 1.4 & 1.1 & 1.1 & 1.0  & 1.1 & 0.6 & 1.4 & \textbf{0.5} & 1.8 & 1.5 & 1.7 & 2.0\\
    &   $\times$ 0.1 & 1.3 &  1.0 & 1.2  & 1.1  & 1.0 & 0.9 & 1.0 & 1.3 & 1.0  & 1.1 & 1.0 & 1.0 & 1.0 & 1.0 & 1.0 & 1.0 & 1.1 & 1.1 & 1.0 & 0.9 & 1.0\\
\hline
 ${^{48}}{\rm Cr}(\alpha,p){^{51}}{\rm Mn}$ & $\times$ 100 & 0.6 & 0.8 & \textbf{0.4} &   0.7 & 0.7 & 0.9  & 1.0 &  1.6 & 1.2 & 1.5 & 1.2 & 1.1 & 1.1 & 1.1 & \textbf{0.3} & 0.7 & \textbf{0.2} & 0.6 & 0.6 & 0.8 & 0.9\\
    &   $\times$ 0.01 & 1.2 & 1.1  & 1.1  & 1.0 & 1.1 & 1.0 & 1.0 &  1.4 & 1.0  & 0.9  & 1.0 & 1.0 & 1.0 & 1.0 & 0.9 & 1.1 & 1.2 & 1.0  & 1.2 & 1.0 & 1.0 \\
\hline
${^{52}}{\rm Fe}(\alpha,p){^{55}}{\rm Co}$ & $\times$ 100 & 0.7 & \textbf{2.8} & \textbf{0.5} & 1.1   & 1.0 & 1.3  & 1.0 &  \textbf{2.1} & 1.1 & 1.6 & 1.2  & 1.1 & 1.1  & 1.1  & \textbf{0.3} & \textbf{2.4} & \textbf{0.3} & 1.0  & 0.9 & 1.2  & 0.9\\
    &   $\times$ 0.01 & 1.2 & 1.0   & 1.0  & 1.0  & 0.9 & 1.1 & 1.0 & 1.3 & 1.0  & 1.3  & 1.0 & 1.0 & 1.0 & 1.0 & 0.9 & 1.0  & 0.8  & 1.0 & 0.9 & 1.1 & 1.0 \\
\hline
${^{54}}{\rm Fe}(\alpha,p){^{57}}{\rm Co}$ &  $\times$ 1.08 & 1.1 & 1.0  & 0.8 & 1.0  & 1.0 & 1.1  & 1.0 &   1.3 & 1.0 & 0.8 & 1.0   & 1.0 & 1.0 & 1.0 & 0.8 & 1.0 & 1.0 & 0.9  & 1.0 & 1.1  & 1.0\\
    & $\times$  0.92 & 1.1 & 1.0   & 0.8  & 1.0  & 0.9  & 0.9 & \textbf{3.6} & 1.2  & 1.0 & 0.7  & 1.0  & 1.0 & 1.0  & 1.0 & 0.9  & 1.0 & 1.2  &1.0  & 1.0  & 1.0 & \textbf{3.6} \\
\hline
${^{56}}{\rm Ni}(\alpha,p){^{59}}{\rm Cu}$ & $\times$ 10 & 0.8  & \textbf{0.5} & 0.6  &   0.8 & 0.8 & 1.0   & 1.0 &  \textbf{2.2} & 1.0 & 1.4 & 1.0  & 1.1 & 1.0 & 1.0 & \textbf{0.4} & \textbf{0.5} & \textbf{0.4} & 0.8  & 0.7 & 1.0 & 1.0\\
    &   $\times$ 0.1 & 1.2 & 1.1  & 1.3   & 1.2  & 1.2 & 1.1 & 1.0 & 1.5  & 1.0  & 1.4  & 1.0 & 1.0 & 1.0 & 1.0 & 0.8 & 1.1  & 0.9 & 1.2  & 1.3 & 1.1  & 1.0 \\
\hline
${^{13}}{\rm N}(\alpha,p){^{16}}{\rm O}$ & $\times$ 100 & 0.8  & 1.1 & \textbf{0.5} & 1.0 & 0.8 & 0.6   & 0.9 &  0.7 & \textbf{0.4} & \textbf{0.1} & \textbf{0.5}   & \textbf{0.5}  & 0.6  & 0.7  & 1.0 & \textbf{2.6} & \textbf{3.6} & 1.8  & 1.6 & 1.0 & 1.3 \\
    &   $\times$ 0.01 & 1.3 & 1.9   & 0.9  & 1.4  & 1.4 & 1.6 & 1.1 & 1.7 & 1.2  & 1.3 & 1.0 & 1.1 & 1.0 & 1.0 & 0.8 & 1.6 & 0.7 & 1.4  & 1.3 & 1.6 & 1.0 \\
\hline
${^{43}}{\rm Sc}(\alpha,p){^{46}}{\rm Ti}$ & $\times$ 10 & 1.2 & \textbf{2.1} & 1.0 & 1.1  & 1.1 & 1.2  & 1.0 & 1.5 & 1.1 & 1.1 & 1.0  & 1.0 & 1.0 & 1.0 & 0.8 & 1.9 & 0.9 &  1.1 & 1.1 & 1.2 & 1.0  \\
    &   $\times$ 0.1 & 1.1 & 1.0   & 0.9 & 1.2  & 1.0  & 1.0 & 1.7 & 1.4 & 1.0  & 1.0 & 1.0  & 1.0 & 1.0 & 1.0 & 0.8 & 1.0  & 0.9  &  1.2 & 1.0 & 1.0  & 1.6 \\
\hline
${^{57}}{\rm Ni}(n,p){^{57}}{\rm Co}$ & $\times$ 100 & \textbf{4.0} & \textbf{4.2} & \textbf{4.2} & \textbf{4.8}  & \textbf{3.1} & 1.8  & 1.1 & \textbf{0.3} & 0.8 & \textbf{0.4} & 0.8 & 0.7  & 0.9  & 0.8 & \textbf{13.4} & \textbf{5.1} & \textbf{10.0} & \textbf{5.7}  & \textbf{4.3} & \textbf{2.1}  & 1.4\\
    &   $\times$ 0.01 & 0.7 & 0.6   & \textbf{0.5} & 1.1  & 0.8 & 1.0 & 1.0 & 1.8 & 1.1  & 1.6 & 1.1 &1.0 & 1.0 & 1.0 & \textbf{0.4} & \textbf{0.5}  & \textbf{0.3}  & 1.1 & 0.8 & 1.0 & 1.0 \\
\hline
${^{56}}{\rm Co}(p,n){^{56}}{\rm Ni}$ & $\times$ 100 & \textbf{0.5}  & 1.0 & \textbf{0.3} & 1.0   &  0.8 & 0.9 & 1.0 & \textbf{2.4} & 1.2 & 1.8 & 1.3 & 1.1 & 1.1  & 1.1 &  \textbf{0.2} & 0.8 &  \textbf{0.2} &  0.8 & 0.7 & 0.8 & 0.9 \\
    &   $\times$ 0.01 & 1.6 & 1.8   & 1.6  & \textbf{2.0} & 1.6 & 1.2 & 1.0  & 1.2 & 0.9  & 1.0  & 0.9 & 0.9 & 0.9 & 0.9 & 1.4 & \textbf{2.0}  & 1.7  & \textbf{2.3}  & 1.8 & 1.3 & 1.1 \\
\hline

${^{47}}{\rm V}(p,\gamma){^{48}}{\rm Cr}$ & $\times$ 10 & 0.7 & \textbf{0.5} & \textbf{0.5} & \textbf{0.5} &  0.7 & 0.9  & 1.0  & \textbf{2.1}  & 1.1 & 1.4 & 1.1  & 1.1 & 1.0  & 1.1 & \textbf{0.3} & \textbf{0.4} &  \textbf{0.4} & \textbf{0.5}  & 0.6 & 0.9  & 0.9 \\
    &   $\times$ 0.1 & \textbf{2.0} & \textbf{2.4} & \textbf{2.0} &  \textbf{2.2} & \textbf{2.0} & 1.5 & 1.1  & 1.0 & 1.0 & 0.7  & 1.0  &  0.9 & 1.0 & 0.9 & \textbf{2.2} & \textbf{2.5}  & \textbf{2.7} & \textbf{2.3}  & \textbf{2.2} & 1.6 & 1.1  \\
\hline
${^{42}}{\rm Ca}(p,\gamma){^{43}}{\rm Sc}$ & $\times$ 1.21 & 1.1  & 0.9 & 1.0 & 1.0  & 1.0 & 0.9 & 1.0 & 1.4 & 1.1  & 1.2 & 1.0 & 1.0  & 1.0  & 1.0 & 0.8 & 0.8 & 0.9 &  1.0 & 1.0  & 0.9  & 1.0 \\
    & $\times$  0.79 & 1.1  & 1.0   & 1.0 & 1.1 & 0.9 & \textbf{2.9} & 1.0 & 1.2 & 1.0 & 1.1 & 1.0 & 1.0 & 1.0 & 1.0 & 0.9 & 1.0 & 0.9  & 1.2  &  0.9 & \textbf{2.9}  & 1.0 \\
\hline
${^{39}}{\rm K}(p,\gamma){^{40}}{\rm Ca}$ & $\times$ 100 & \textbf{5.2} & \textbf{3.5} & \textbf{4.0} & \textbf{3.5} & \textbf{3.4} & \textbf{2.4}   & 1.2 & 0.6 & 1.0 &  1.1 & 1.0  &1.0  & 1.0 & 1.0 & \textbf{9.2} & \textbf{3.5} & \textbf{3.5} & \textbf{3.4}  & \textbf{3.3} & \textbf{2.4} & 1.2 \\
    &   $\times$ 0.01 & \textbf{0.1} & \textbf{0.2}   & \textbf{0.1}  & \textbf{0.2}  & \textbf{0.2} & \textbf{0.5} & 0.9 & 1.1 & 1.0  & 0.8  & 0.9  & 0.9 & 1.0 & 1.0 & \textbf{0.1} & \textbf{0.2} & \textbf{0.2}  & \textbf{0.2}  & \textbf{0.2} & \textbf{0.5} & 0.9 \\
\hline
${^{57}}{\rm Co}(p,\gamma){^{58}}{\rm Ni}$ & $\times$ 10 & \textbf{2.1} & \textbf{2.3} & 1.9  & \textbf{2.0}  & 1.5 & 1.1  & 1.1  & 1.2 & 0.9 & 0.9 & 0.9  & 0.9 & 1.0  & 0.9 & 1.8 & \textbf{2.5} & \textbf{2.2} & \textbf{2.2}  & 1.7 &  1.2 & 1.2\\
    &   $\times$ 0.1 & 1.2  & 0.6   & 0.9  & 0.8 & 0.8 & 0.8 & 1.0 & 1.9 & 1.1  & 1.1  &  1.0 & 1.0 & 1.0 & 1.0 & 0.6 & 0.6  & 0.8  &  0.7 & 0.8 & 0.8  & 0.9 \\
\hline
${^{52}}{\rm Mn}(p,\gamma){^{53}}{\rm Fe}$ & $\times$ 10 & 0.7 & 0.6  & \textbf{0.5}   & 0.9 & 0.9 & 1.1  & 1.0 & 1.8  & 1.1 & 1.2 & 1.0  & 1.1 & 1.0  & 1.1 & \textbf{0.4} & \textbf{0.5} & \textbf{0.5} & 0.9  & 0.8 & 1.0  & 0.9 \\
    &   $\times$ 0.1 & 1.5 & 1.4   & 1.6   & 1.4 & 1.3 & 1.3 & 1.0  & 0.9 & 0.9  & 0.9  & 0.9 & 0.9 & 0.9 & 0.9 &  1.6 & 1.5  & 1.9  & 1.5  & 1.4 & 1.4 & 1.1 \\
\hline
${^{27}}{\rm Al}(\alpha,n){^{30}}{\rm P}$ & $\times$ 10 & 1.0 & 1.0 & 1.0 & 1.0  & \textbf{2.9} & 0.9  & 1.3  &  1.2 &1.0  & 0.9 & 1.0  & 1.0 &  1.0 & 1.0 & 0.9 & 1.0 & 1.1 & 1.0   & \textbf{3.0}  & 0.9 & 1.4\\
    &   $\times$ 0.1 & 1.2 & 1.1   & 1.1  & 0.9 & 1.1 & 1.0 & 1.0 & 1.2 & 1.0  & 1.4 & 1.0 & 1.0 & 1.0 & 1.0 & 1.0 & 1.1 & 0.8  & 0.9 &  1.1 & 1.0 & 1.0 \\
\hline
${^{39}}{\rm K}(p,\alpha){^{36}}{\rm Ar}$ & $\times$ 100 & \textbf{0.1} & \textbf{0.2} & \textbf{0.1} & \textbf{0.3}  & \textbf{0.3} & 0.6  & 0.7 & 0.9 & 0.8 & 0.6 &0.9   & 0.8 & 0.9  &0.9  & \textbf{0.1} & \textbf{0.3} & \textbf{0.2} &\textbf{0.3}   & \textbf{0.3} & 0.6  & 0.8 \\
    &   $\times$ 0.01 & \textbf{3.3} & \textbf{2.6}   & \textbf{2.4}  & \textbf{2.6}  & \textbf{2.4} & \textbf{2.0} & 1.2 & 1.0 & 1.0 & 1.2 &1.0  & 1.0 & 1.0 & 1.0 & \textbf{3.4} & \textbf{2.6}  &  \textbf{2.1} & \textbf{2.6}  & \textbf{2.4} & \textbf{2.0} & 1.2 \\
\hline
 
  
\end{tabular}

\end{longrotatetable}
\egroup

\subsection{Reactions impacting ${^{56}}$\rm{Ni}} \label{sec:ni56}

Here we discuss the reactions listed in Table~\ref{tab:sen_val} that significantly impact $M(^{56}{\rm Ni})$. As highlighted by \citet{Magkotsios_2010}, $^{56}{\rm Ni}$ nucleosynthesis is far less sensitive to nuclear reaction rate variations than $^{44}{\rm Ti}$ because it is much nearer to the global minimum in the nuclear binding energy surface. 
We find sensitivities for nuclear reaction rates impacting the flow into the large QSE cluster  at early times and into the small QSE cluster present around $^{56}{\rm Ni}$
at late times.

For $\alpha$-rich freeze-out conditions, the $3\alpha$$\rightarrow$$^{12}{\rm C}$ reaction contributes flow to the single large QSE cluster that is initially present, driving the formation of the cluster QSE-1~\citep{Magkotsios_2010}. $^{13}{\rm N}(\alpha,p)^{16}{\rm O}$, $^{17}{\rm F}(\alpha,p)^{20}{\rm Ne}$, and $^{27}{\rm Al}(\alpha,p)^{30}{\rm Si}$ all moderate flow from the QSE-1 cluster to higher-mass nuclides. This has a significant impact on $M(^{56}{\rm Ni})$ for a subset of our model calculations. For each of these reactions, an increase in the reaction rate decreases $^{56}{\rm Ni}$ yields.

The ${^{57}}{\rm Ni}(n,p){^{57}}{\rm Co}$ reaction drives material away from the nickel isotopes towards lower mass nuclides when the large QSE cluster in this region has dissolved. Therefore, increasing this reaction rate decreases $M(^{56}{\rm Ni})$ for a subset of the model conditions.


\subsection{Reactions impacting $^{44}{\rm Ti}/^{56}{\rm Ni}$} \label{sec:ti44_ni56}
Inspecting Table~\ref{tab:sen_val} reveals several reactions which impact $M(^{44}{\rm Ti})$ and/or $M(^{56}{\rm Ni})$ for some set of model conditions, but not above our pre-specified significance threshold. However, as the impacts on the two yields are often anti-correlated, two non-significant sensitivities can combine to form a significant impact on the ratio $M(^{44}{\rm Ti})/M(^{56}{\rm Ni})$. As this ratio is considered a distinct diagnostic for model-observation comparisons~\citep[e.g.][]{2017ApJ...842...13W}, we separately highlight these reaction rate sensitivities in Table~\ref{tab:sen_val}.


\section{Discussion} \label{sec:comp}

It is interesting to compare the results of the present work to previous studies exploring nuclear reaction rate sensitivities of shock-driven nucleosynthesis. In particular, we focus our discussion on the post-processing studies of \citet{The_1998} and \citet{Magkotsios_2010}, as these studies also explored the impact of individual reaction rate variations for a large number of rates. The pioneering work of \citet{The_1998} used an exponential temperature and density trajectory (see Equation~\ref{eqn:adtemp}) with $T_{0}=5.5$~GK, $\rho_{0}=10^{7}$~g\,cm$^{-3}$, $Y_{e}=0.497~\rm{and}~0.499$, and reaction rate variation factors of 100. \citet{Magkotsios_2010} expanded on this work, exploring reaction rate sensitivities with rate variation factors of 100 for exponential and power law trajectories (see Equation~\ref{eqn:pow}), a large phase space of $T_{0}$--$\rho_{0}$ (see Figure~\ref{fig:temp_dens}), and $Y_{e}=0.48-0.52$. In comparison, our model calculations have average $Y_{e}=0.498-0.499$, thermodynamic trajectories more closely resembling the power-law forms, with mass zones covering the $T_{0}$--$\rho_{0}$ phase-space as shown in Figure~\ref{fig:temp_dens}, and focused mostly on the strong reactions identified as important by \citet{Magkotsios_2010} using rate variation factors informed by the literature.

While our list of influential reactions shares several similarities with previous works, there are some conspicuous differences. In particular, we do not highlight many of the reactions that both prior studies identified as high-impact\footnote{Here, we consider Table~4 of \citet{The_1998} and reactions designated as ``primary" in Table~3 of \citet{Magkotsios_2010}.}, namely, $^{40}{\rm Ca}(\alpha,\gamma)^{44}{\rm Ti}$, $^{44}{\rm Ti}(\alpha,p)^{47}{\rm V}$, $^{45}{\rm V}(p,\gamma)^{46}{\rm Cr}$, and $^{57}{\rm Ni}(p,\gamma)^{58}{\rm Cu}$. For each of these reactions, the likely explanation is that we used far smaller rate variation factors (see Table~\ref{tab:table_rr}). For the first two in this set, this is based on the significant efforts in the nuclear physics community to reduce these reaction rate uncertainties. For the latter two, our reduced rate variation factors (10 as opposed to 100) are based on the presumed accuracy of Hauser-Feshbach reaction rate predictions given the high nuclear level densities involved. 

Our relative insensitivity to variations in the $^{40}{\rm Ca}(\alpha,\gamma)^{44}{\rm Ti}$ and $^{44}{\rm Ti}(\alpha,p)^{47}{\rm V}$ rates when using experimentally constrained  uncertainties\footnote{We performed test calculations using a rate variation factor of 100 for these two rates with $M_{\rm prog}=15~M_{\odot}$ $E_{\rm inj}=1.32$~foe, and found $M(^{44}{\rm Ti})$ variations similar to Figure~14 of \citet{Magkotsios_2010}.} is in agreement with \citet{Hoff10} who sampled a slightly lower $\rho_{0}$ region of the thermodynamic phase space with models of CasA from \citet{2008nuco.confE.112M}. They found that remaining uncertainties in $^{40}{\rm Ca}(\alpha,\gamma)^{44}{\rm Ti}$ contribute a $\sim20$\% variation to $X(^{44}{\rm Ti})$, while $^{44}{\rm Ti}(\alpha,p)^{47}{\rm V}$ uncertainties (estimated there as a factor of 3) lead to a 70\% variation. Similarly, \citet{Tur_2010} found the $3\alpha$$\rightarrow$$^{12}{\rm C}$ and $^{12}{\rm C}(\alpha,\gamma)^{16}{\rm O}$ reactions are sufficiently constrained from the standpoint of $^{44}{\rm Ti}$ CCSN nucleosynthesis.

For the remaining high-impact cases of \citet{Magkotsios_2010}, the sensitivity identified in their work is for a region of phase space that our models did not populate. For $^{17}{\rm F}(\alpha,p)^{20}{\rm Ne}$, $^{21}{\rm Na}(\alpha,p)^{24}{\rm Mg}$, and $^{40}{\rm Ca}(\alpha,p)^{43}{\rm Sc}$, \citet{Magkotsios_2010} only find a significant impact for relatively proton-rich initial conditions. For $^{41}{\rm Sc}(p,\gamma)^{42}{\rm Ti}$ and $^{44}{\rm Ti}(p,\gamma)^{45}{\rm V}$, \citet{Magkotsios_2010} observe significant variations of $X(^{44}{\rm Ti})$ only for quite high-$T_{0}$ low-$\rho_{0}$ conditions.

Indeed the significant reaction rate sensitivities that we identify in Table~\ref{tab:table_rr} are all highlighted as of secondary importance by \citet{Magkotsios_2010}. However, since their ``secondary" designation is a general term for less than a factor of 10 variation in $X(^{44}{\rm Ti})$ for any region within their full $T_{0}$--$\rho_{0}$ phase space, a quantitative comparison is not possible. Qualitatively, we find a significant sensitivity in $M(^{44}{\rm Ti})$ and/or $M(^{44}{\rm Ti})/M(^{56}{\rm Ni})$ for the majority of $(\alpha,p)$ and $(p,\gamma)$ reactions identified by \citet{Magkotsios_2010} as impacting the chasm depth, and a subset of the $(p,n)$ reactions found to impact the chasm width. The connection of our results to the chasm are not surprising, as all of our model calculations have mass zones very near to or directly over this region.

Perhaps more interesting are the reaction rate sensitivities that we find in our work which were not identified by \citet{Magkotsios_2010}. We included $^{39}{\rm K}(p,\alpha)^{36}{\rm Ar}$ and $^{27}{\rm Al}(\alpha,n)^{30}{\rm P}$ as a test of the completeness of the previous survey, finding significant sensitivities for some choices of model conditions. This is not entirely unexpected, as \citet{The_1998} identified $^{27}{\rm Al}(\alpha,n)^{30}{\rm P}$ and $^{36}{\rm Ar}(\alpha,p)^{39}{\rm K}$, the reverse of $^{39}{\rm K}(p,\alpha)^{36}{\rm Ar}$, as influential for some choices of $Y_{e}$. Nonetheless, this highlights the need for larger-scale reaction rate sensitivity studies for shock-driven nucleosynthesis using one-dimensional models.

\section{Conclusions} \label{sec:conc}

We performed nuclear reaction rate sensitivity studies for $^{44}{\rm Ti}$ and $^{56}{\rm Ni}$ production in CCSN shock-driven nucleosynthesis using the code {\tt MESA}. We evolved a range of $M_{\rm prog}$ stars to core collapse and induced an artifical explosion with a range of $E_{\rm inj}$, analyzing nucleosynthesis yields for a range of $M_{\rm cut}$, in order to gauge the robustness of our results to model calculation assumptions. For each set of model assumptions, we varied the strong reactions previously identified by \citet{Magkotsios_2010} as influential for $^{44}{\rm Ti}$ production, including two additional rates to test the completeness of the influential reaction rate set, using reaction rate variation factors based on uncertainty constraints in the literature. 

We find a significant impact on $^{44}{\rm Ti}$ and/or $^{56}{\rm Ni}$ nucleosynthesis for only a subset of the influential reaction rates of \citet{Magkotsios_2010}, though this is likely attributable to the smaller reaction rate variation factors and different astrophysical conditions used in our work. Additionally, we find nucleosynthesis sensitivities for reaction rates not identified by \citet{Magkotsios_2010}. This indicates that larger-scale reaction rate sensitivity studies exploring an expanded set of reaction rates is desirable. 

From the present work we conclude that additional effort is required from the nuclear physics community to reduced the reaction rate uncertainties for $^{13}{\rm N}(\alpha,p)^{16}{\rm O}$, $^{17}{\rm F}(\alpha,p)^{20}{\rm Ne}$, $^{27}{\rm Al}(\alpha,p)^{30}{\rm Si}$, $^{43}{\rm Sc}(\alpha,p)^{46}{\rm Ti}$, $^{48}{\rm Cr}(\alpha,p)^{51}{\rm Mn}$, $^{52}{\rm Fe}(\alpha,p)^{55}{\rm Co}$, $^{54}{\rm Fe}(\alpha,p)^{57}{\rm Co}$, $^{55}{\rm Co}(\alpha,p)^{58}{\rm Ni}$, $^{56}{\rm Ni}(\alpha,p)^{59}{\rm Cu}$, $^{57}{\rm Ni}(n,p)^{57}{\rm Co}$, $^{56}{\rm Co}(p,n)^{56}{\rm Ni}$, $^{39}{\rm K}(p,\gamma)^{40}{\rm Ca}$, $^{42}{\rm Ca}(p,\gamma)^{43}{\rm Sc}$,  $^{47}{\rm V}(p,\gamma)^{48}{\rm Cr}$, $^{52}{\rm Mn}(p,\gamma)^{53}{\rm Fe}$, $^{57}{\rm Co}(p,\gamma)^{58}{\rm Ni}$,  $^{27}{\rm Al}(\alpha,n)^{30}{\rm P}$, and $^{39}{\rm K}(p,\alpha)^{36}{\rm Ar}$.

\acknowledgements
We thank the Ohio Supercomputer Center for providing computational resources~\citep{OhioSupercomputerCenter1987}, R. Farmer for making his pre-supernova evolution inputs for {\tt MESA} available and for assistance in their implementation,  and F.X. Timmes for sharing the data from \cite{Magkotsios_2010}.
This work was supported by the U.S. Department of Energy under grants DE-FG02-88ER40387 and DE-SC0019042, the National Nuclear Security Administration under grant DE-NA0003909, and has benefited from
support by the National Science Foundation under grant PHY-1430152 (Joint Institute for Nuclear Astrophysics--Center for
the Evolution of the Elements).

{\it Software:} Modules for Experiments in Stellar Astrophysics
({\tt MESA})\citep{2011ApJS..192....3P,Paxt13,2015ApJS..220...15P,2018ApJS..234...34P}

\bibliographystyle{aasjournal}
\bibliography{ti44_ni56_sen_stdy}

\end{document}